\documentclass[showpacs,aps,prd,reprint,superscriptaddress,nofootinbib,floatfix]{revtex4-1}
\usepackage[dvipdfmx]{graphicx}

\usepackage{amsmath,amssymb,bm,color,longtable,mathrsfs}%,slashed}
\usepackage{ulem}
\usepackage{multirow,bigdelim}

\usepackage{hyperref}
\hypersetup{
    colorlinks,
    linkcolor={red!50!black},
    citecolor={blue!50!black},
    urlcolor={blue!80!black}
}

\usepackage{footmisc}

\usepackage{xcolor}
\def\be#1{\begin{equation}#1\end{equation}} %{equation}

\def\beqnn#1{\begin{eqnarray}#1\end{eqnarray}}
\begin{document}
 \begin{flushright}
% %KEK-TH-????
KANAZAWA-26-07
 \end{flushright}

\title{
All-mode approach for general tensor renormalization group method
}

\author{Katsumasa~Nakayama}
\email[]{katsumasa.nakayama@riken.jp}
\affiliation{RIKEN Center for Computational Science, Kobe, 650-0047, Japan}
\author{Shinji~Takeda}
\email[]{takeda@hep.s.kanazawa-u.ac.jp}
\affiliation{Institute for Theoretical Physics, Kanazawa University, Kanazawa 920-1192, Japan}

\begin{abstract}
We apply the stochastic all-mode technique to general coarse-graining schemes of the tensor renormalization group (TRG) approach.
The technique possesses a notable characteristic of being entirely free from systematic errors which can potentially arise in the coarse-graining step, while it has statistical errors caused by random noise.
The key element in our generalization is employing a method called squeezer, and the all-mode technique can be used when it is constructed.
Since many TRG methods can be described in terms of the squeezers, our approach can be applicable in all such cases.
In this work, we demonstrate the new method with higher-order TRG in the two- and three-dimensional Ising model.

\end{abstract}


\maketitle

%\tableofcontents

\section{Introduction}
Tensor networks are widely used in the field of physics and have attracted particular attention. 
In recent years, they have shown remarkable power in numerical computations.
A striking feature of the tensor network methods is employing information compression techniques based on singular value decomposition (SVD), while importance sampling is usually not used in contraction calculations.
This characteristic makes tensor network approaches suitable for computing physical quantities without encountering the notorious sign problem in the Monte Carlo method.
For example, the density matrix renormalization group \cite{DMRG} using matrix product states \cite{DMRG_MPS}, projected entangled pair states \cite{PEPS}, and %as a distinct method, 
the tensor renormalization group (TRG) \cite{Levin:2006jai} have successfully computed various physical quantities, even for critical phenomena \cite{Z2andmore,Kuramashi:2018mmi,SchwingerOneFlavor,SchwingerTopological,SchwingerPhaseTransition,Yu:2013sbi,Zou:2014rha,Yang:2015rra,GrossNeveu,Grassmann2018,Bazavov:2019qih,U1theta,CP1TRG,Hirasawa2021,Akiyama2021,TRG_MultiFlavor,3dSU2chiral,SU2SU3}.

As an approximation technique for tensor network contractions, the TRG has proven to be a powerful method and has undergone significant development in the past two decades.
For example, the original TRG method has been improved by using different singular value weighting schemes~\cite{Levin:2006jai,bwTRG}, the truncated singular value method ~\cite{rand_trunc,Nakamura:2018enp,RandTRG,redsvd}, and variational approaches \cite{Evenbly2015,LoopTNR,Entbranch}, etc.
Furthermore, while the original TRG was limited to two-dimensional systems, it was later extended to systems with more than three dimensions, which is referred to as higher-order TRG (HOTRG) \cite{HOTRG}. The computational cost of HOTRG for higher dimensional system was prohibitively high, however, algorithms addressing this problem were subsequently proposed \cite{ATRG,TriadRG,MDTRG}.
%Building on these TRG advancements, it is now possible to compute physical quantities such as free energy and the expectation values of local operators.
In this way, feasibility for higher-dimensional systems was to some extent established, and many higher-dimensional calculations began to be carried out \cite{
Akiyama:2019xzy,
Akiyama:2021zhf,
Akiyama:2022eip,
Akiyama:2023hvt,
Sugimoto:2025vui,
Sugimoto:2025xva,
Sugimoto:2026wnw}.

Although TRG-based methods have achieved significant progress in recent years, the assessment of systematic errors remains a major challenge. In principle, one can estimate the truncation error at each coarse-graining step by monitoring the discarded singular values in the SVD. However, evaluating the total systematic error in physical observables is far from straightforward, as the propagation and accumulation of truncation errors through the iterative coarse-graining procedure are not yet fully understood.
Assuming that the systematic error decreases simply with increasing bond dimension, a practical approach is to compute physical quantities for several bond dimensions and extrapolate the results using an appropriate functional form. Alternatively, one may estimate the systematic uncertainty from the variation of the results with respect to the bond dimension. Nevertheless, establishing a reliable and systematic error estimation scheme remains one of the central challenges of TRG methods and is essential for achieving high-precision calculations in the future.

One approach to this problem was originally proposed by Ferris \cite{ferris2015unbiasedmontecarloage}, who introduced the idea of incorporating Monte Carlo sampling into tensor network methods. This framework was subsequently developed into a practical implementation by the Berkeley group \cite{huggins2017montecarlotensornetwork}, and has recently been revived and further refined by Todo \cite{2024arXiv241202974T}.
The basic idea is introducing the stochastic sampling in order to remove the systematic error due the truncation.
Another statistical approach was given in \cite{Allmode} by Ohki et al., so called all-mode method \cite{Foley_2005} which is originally proposed in lattice QCD calculations.
In the \cite{Allmode}, the authors incorporate the all-mode technique into the original TRG algorithm, and they successfully reproduce exact results of two-dimensional Ising model up to volume $V=2^{12}$.

In this paper, we aim to generalize the all-mode technique to a wider class of tensor-network algorithms beyond TRG, especially to higher-dimensional formulations such as HOTRG \cite{HOTRG}, with prospective extensions to anisotropic TRG \cite{ATRG} and triad TRG \cite{TriadRG}.
In such algorithms, the information compression processes can be basically described by squeezers \cite{ATRG}, which were introduced in the context of the boundary TRG method \cite{boundaryHOTRG}.
We have found that the all-mode technique is highly compatible with constructing the squeezer.
Therefore, our approach enables the application of the all-mode technique to systems beyond two dimensions.

This paper is organized as follows.
In Section \ref{AM_squeezer}, we introduce the squeezer and the all-mode technique.
Building on this foundation, Section \ref{AM_HOTRG} presents its application to the two-dimensional HOTRG and extends the approach to three dimensions.
In Section \ref{result}, we demonstrate the effectiveness of the proposed method by applying it to the Ising model in both two- and three-dimensional systems.
Finally, the conclusions are summarized in Section \ref{concl}.

%%%%%%%%%%%%%%%%%%%%%%%%%%%%%%%%%%%%%%%%%%%
%\section{All-mode approach with the squeezer}
\section{Squeezer and all-mode technique}
\label{AM_squeezer}

To introduce the squeezer, we begin by discussing a decomposition for a general tensor network.
In the following, the bond dimension of a single line is assumed to be $D$.
For a given tensor network ${\cal N}$
which is supposed to have a plaquette (loop) structure, if one wishes to decompose it in a tree-like shape, a singular value decomposition (SVD) may be a natural choice,
\be{\label{eq:TNgeneral}
\begin{aligned}
\vcenter{\hbox{\includegraphics[height=2.4cm]{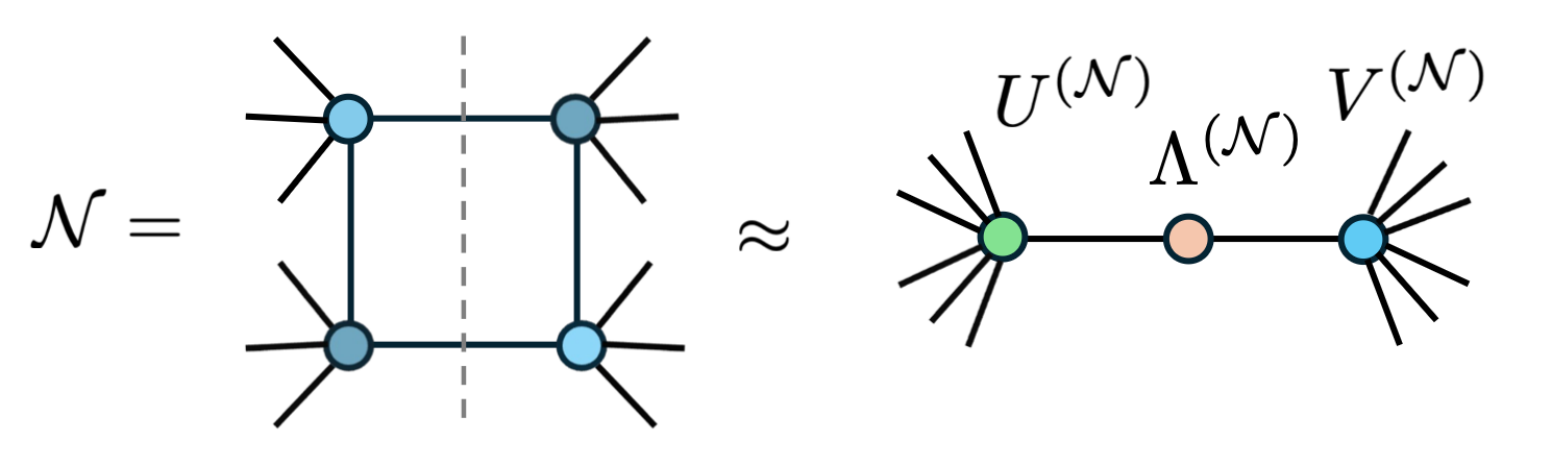}}}
\end{aligned}.
}
However, when the tensor network has many outer indices, performing the SVD in a straightforward manner leads to a huge computational cost.
One way to avoid this issue is to use the squeezer in Ref. \cite{boundaryHOTRG,ATRG} which plays a similar role to the isometry for HOTRG,
$
{\cal N}
=
AB \approx AP_1P_2B$, namely
\be{
\begin{aligned}
\vcenter{\hbox{\includegraphics[height=2.4cm]{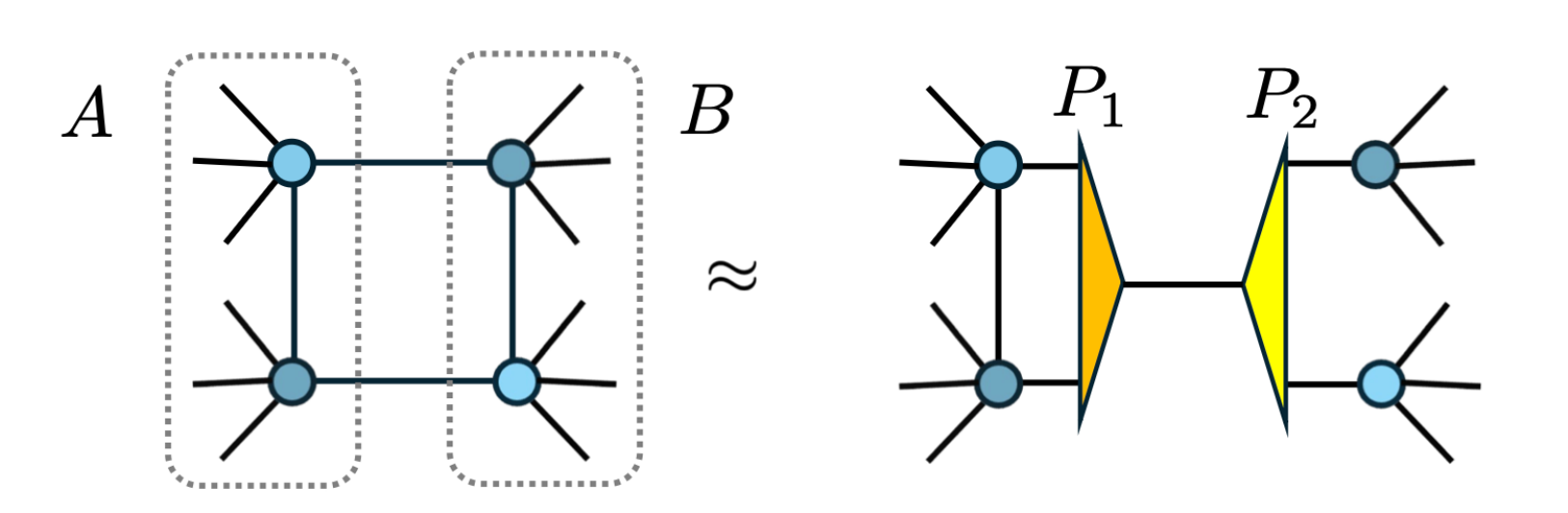}}}
\end{aligned}
\label{eqn:AP1P2B}.
}
The squeezers $P_1$ and $P_2$ are determined by minimizing the following cost function,
\be{
\min_{P_1,P_2}
||AB - AP_1P_2B||_{\rm F}^2.
}

A practical way to determine the squeezers starts from splitting the target tensor network into two parts $A$ and $B$, and then
considering SVD~\footnote{In an actual computation, we do not have to fully compute the SVD of $A$ and $B$. What we only need is the singular values and either of unitary matrices as explained below.} for each sub-network,
\be{
\begin{aligned}
\vcenter{\hbox{\includegraphics[width=6.5cm]{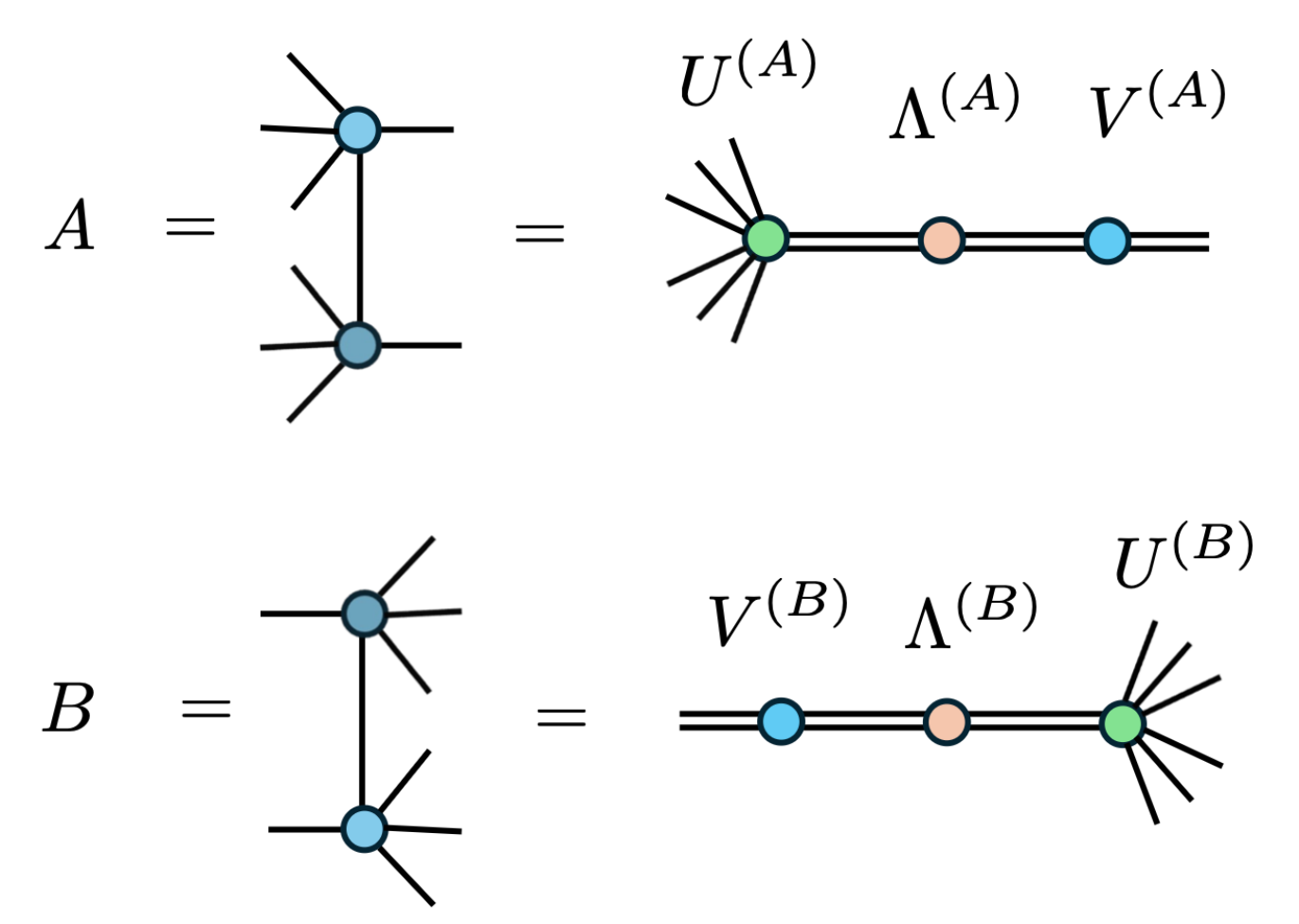}}}
\end{aligned},
\label{eqn:AB_SVD}
}
and then the total network is given by
\be{
{\cal N}
=
AB
=
U^{(A)}\Lambda^{(A)}V^{(A)}
\,\,\,
V^{(B)}\Lambda^{(B)}U^{(B)}.
\label{eqn:ABULVULV}
}
To transform the tensor network above into the structure shown in the right hand side of Eq.~\eqref{eqn:AP1P2B}, we further apply the SVD for the middle part ($D^2\times D^2$ matrix) in Eq.~\eqref{eqn:ABULVULV} which involves two singular values matrices as follows and keep the largest $D$ modes,
\be{
\begin{aligned}
\vcenter{\hbox{\includegraphics[width=7.5cm]{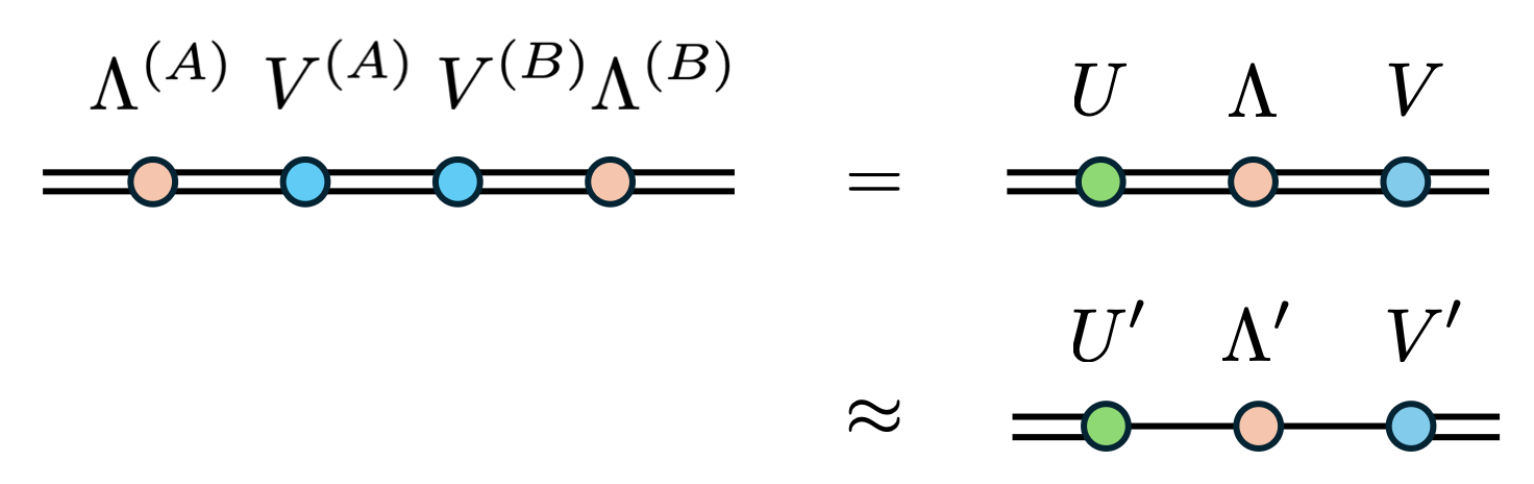}}}
\end{aligned}
\label{eqn:ULV}
}
where note that in the last step the truncation error is introduced, say  $U^{\prime}$ is $D^2\times D$ matrix.
Comparing Eq.~\eqref{eqn:AP1P2B} and Eq.~\eqref{eqn:ABULVULV} using Eq.~\eqref{eqn:ULV}
leads
\beqnn{
AP_1
=
U^{(A)}U^{\prime}\sqrt{\Lambda^{\prime}}
&\Longleftrightarrow&
\Lambda^{(A)}V^{(A)}P_1
=
U^{\prime}\sqrt{\Lambda^{\prime}},
\\
P_2B
=
\sqrt{\Lambda^{\prime}}V^{\prime} U^{(B)}
&\Longleftrightarrow&
P_2 V^{(B)}\Lambda^{(B)}
=
\sqrt{\Lambda^{\prime}}V^{\prime}.
}
From above equations, one can read off $P_1$ ($P_2$) using $V^{(A)}$
($V^{(B)}$),
\beqnn{
P_1
&=&
V^{(A)\dag}\frac{1}{\Lambda^{(A)}}U^{\prime}\sqrt{\Lambda^{\prime}},
\\
P_2
&=&
\sqrt{\Lambda^{\prime}}V^{\prime} \frac{1}{\Lambda^{(B)}}V^{(B)\dag}.
}
Note that $P_1$ ($P_2$) is $D^2\times D$ ($D\times D^2$) matrix.
Practically, we use the following equivalent expression for the squeezers %for $P_1$ and $P_2$,
\beqnn{
P_1
&=&
V^{(B)}\Lambda^{(B)}V^{\prime\dag}/\sqrt{\Lambda^{\prime}},
\\
P_2
&=&
(1/\sqrt{\Lambda^{\prime}})U^{\prime\dag} \Lambda^{(A)}V^{(A)},
}
which is derived by using Eq.~\eqref{eqn:ULV} up to the truncation errors.
An important point here is that the squeezers can be constructed solely from $\Lambda^{(A,B)}$ and $V^{(A,B)}$, which are $D^2\times D^2$ matrices, without requiring $U^{(A,B)}$ carrying more indices as shown in Eq.~\eqref{eqn:AB_SVD}.
Such required ingredients can be obtained from the eigenvalue decomposition of $A^\dag A$ and $B B^\dag$, which are $D^2\times D^2$ matrix
\beqnn{
A^\dag A&=&V^{(A)\dag}(\Lambda^{(A)})^2 V^{(A)},
\\
B B^\dag&=&V^{(B)}(\Lambda^{(B)})^2 V^{(B)\dag}.
}
In fact, the cost to obtain $V^{(A)}$ ($V^{(B)}$) in this way is much cheaper than that of SVD for $A$ ($B$) itself in Eq.~\eqref{eqn:AB_SVD}.

A key step in constructing the squeezers is the truncation of the SVD in Eq.~\eqref{eqn:ULV}. 
In contrast, the construction of the conventional isometry in the standard HOTRG does not involve a truncated SVD but it simply truncates the unitary matrix which does not correspond to a weighted truncation.
In any case, the truncation introduced in Eq.~\eqref{eqn:ULV} becomes a source of nontrivial systematic errors.
To eliminate such errors, here we introduce stochastic noise into the decomposition, following Ref.~\cite{Allmode}.
First, we prepare a noise vector~\footnote{One can generate $N_{\rm r}(\ge1)$ independent noise vectors to reduce statistical noise. In practice, however, we find that increasing $N_{\rm r}$ does not significantly improve the final precision. Therefore, throughout this paper, we set $N_{\rm r}=1$ exclusively. } $\eta$ whose dimensionality is $D^2 - (D-1)$.
For simplicity, we adopt $Z_2$ noise, i.e., $\eta_{i}\in \{-1,+1\}$ for $i=1,2,\dots,D^2-D+1$.
This choice has an important property: if we generate $N_\mathrm{stat}$ independent noise vectors $\eta^{[\ell]}$ ($\ell=1,2,\ldots,N_{\rm stat}$) and take an ensemble average, they satisfy an approximate completeness relation
\beqnn{
\langle
\eta\eta^\dag
\rangle
&=&
\frac{1}{N_\mathrm{stat}}
\sum_{\ell = 1} ^{N_\mathrm{stat}}
\eta^{[\ell]}
\eta^{[\ell]\dagger}
\nonumber\\
&=&
\mathbf{1}_{(D^2-D+1)\times(D^2-D+1)}
+
O\left(\frac{1}{\sqrt{N_\mathrm{stat}}}\right),
}
where $\langle ... \rangle$ denotes the ensemble average and  the diagonal elements are exactly unity, while the off-diagonal elements contain statistical noise which is suppressed in the limit $N_{\rm stat}\to\infty$.

To see how the noise vector is used in the decomposition, we define the $D^2\times D$  matrix $\xi$, for a given $\ell$
\beqnn{
\xi_{ij}^{[\ell]}
&=&
\begin{cases}
1
\mathrm{\ \ \ }&(i=j=1,2,\ldots,D-1)\\
\eta_{i-D+1}^{[\ell]}\mathrm{\ \ \ }&(i=D,D+1,\ldots,D^2 \mathrm{\ and \ } j=D)\\
0\mathrm{\ \ \ }&(\mathrm{Others})
\end{cases}
\nonumber\\
&=&
\begin{array}{r ccc|c l l}
&\multicolumn{3}{c}{\overbrace{\hspace{4em}}^{D-1}}&\overbrace{}^{1}&&\\
\ldelim[{8}{1pt}&\,\,1& & &0 & \rdelim]{8}{1pt}	&\rdelim\}{3}{1pt}[\scriptsize{$D-1$}]\\
    &&\ddots&&\vdots&&\\
    &&&1&0&&\\
    \hline
    &&&&\eta_1^{[\ell]}&&\rdelim\}{4}{1pt}[\scriptsize{$D^2-D+1$}]\\
    &&\mathbf{0}&&\vdots&&\\
    &&&&\eta_{D^2-D+1}^{[\ell]} &&\\
\end{array}
\nonumber\\
&=&
\begin{aligned}
\vcenter{\hbox{\includegraphics[width=3.cm]{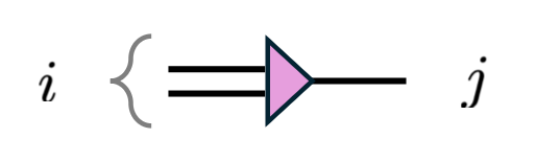}}}
\end{aligned},
}
where the noise vector $\eta$ is embedded. 
Thanks to the completeness of $\eta$,
the matrix $\xi$ follows
\beqnn{
\langle
\xi\xi^\dag
\rangle
&=&
\frac{1}{N_\mathrm{stat}}
\sum_{\ell = 1} ^{N_\mathrm{stat}}
\begin{aligned}
\vcenter{\hbox{\includegraphics[width=2.cm]{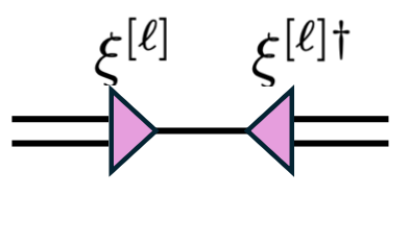}}}
\end{aligned}
\\
&=&
{\bf 1}_{D^2\times D^2}
+
\begin{array}{r ccc|ccc l l}
&\multicolumn{3}{c}{\overbrace{\hspace{4em}}^{D-1}}&\multicolumn{3}{c}{\overbrace{\hspace{5em}}^{D^2-D+1}}\\
\ldelim[{8}{1pt}&\,\,0& & & & & &\rdelim]{8}{1pt}	&\rdelim\}{3}{0.3pt}[\scriptsize{$D-1$}]\\
    &&\ddots&&&\mathbf{0}&&&\\
    &&&0& &&&&\\
    \hline
%    &&&&0&\eta\eta&\eta\eta&&\rdelim\}{4}{0.3pt}
    &&&&0&\langle\eta_1\eta_2\rangle&\cdots&&\rdelim\}{4}{0.3pt}    
    [\scriptsize{$D^2-D+1$}]\\
    &&\mathbf{0}&&
%    \eta_1\eta_2\rangle&\ddots&\eta\eta&\\
    \langle\eta_2\eta_1\rangle&\ddots&\ddots&\\
%    &&&&\eta\eta&\eta\eta &0&\\
        &&&&\vdots&\ddots &0&\\
\end{array}
\nonumber}
where note that the off-diagonal elements ($\eta\eta$) in the lower-right %$(D^2-D+1)\times(D^2-D+1)$ 
sub-block are non-zero due to the noise error, which scales with $1/\sqrt{N_{\rm stat}}$.
Inserting the above equation to the full SVD in Eq (\ref{eqn:ULV}), we obtain
\be{
U
\Lambda
V
=
U
\sqrt{\Lambda}
\langle
\xi\xi^\dagger
\rangle
\sqrt{\Lambda}
V
+
O\left(\frac{\Lambda_D}{\sqrt{N_\mathrm{stat}}}\right).
}
This means that all $D^2$ singular values are taken into account, while the contribution from the higher modes ($D,D+1,...,D^2$) are affected by noise.
If one is interested only in performing the SVD and all singular values are already known, the above expression appears to offer no practical advantage and merely introduces additional statistical errors.
However, using the stochastic matrix $\xi$, we may define stochastic squeezers as
\beqnn{
P_1^{[\ell]}
&=&
(V^{(B)}\Lambda^{(B)}V^{\dag}/\sqrt{\Lambda})\,\xi^{[\ell]},
\\
P_2^{[\ell]}
&=&
(\xi^{[\ell]\dag}/\sqrt{\Lambda})U^{\dag} \Lambda^{(A)}V^{(A)},
}
for each $\ell$ and then by taking ensemble average, they satisfy
\be{
{\cal N}=AB
=A
\langle
P_1P_2
\rangle
B
+
O\left(\frac{1}{\sqrt{N_\mathrm{stat}}}\right).
}
The point here is that the stochastic squeezers contain all singular modes while still retaining a compact representation, say $P_1^{[\ell]}$ is a $D^2\times D$ matrix for a given $\ell$.
Therefore, we can employ these squeezers in the boundary HOTRG coarse-graining procedure, which will be explained in the subsequent section, without encountering an exponential increase in computational cost with the bond dimension while still keeping all modes. Consequently, after taking the ensemble average over all noises for a target physical quantity, for instance the partition function, one can obtain the exact result up to stochastic errors.
This constitutes the basic strategy of the present approach.

\section{All-mode approach for HOTRG}
\label{AM_HOTRG}
In this section, we explain how to apply the all-mode technique to (boundary) HOTRG.
We first present the formulation for the two dimensional square lattice and then extend it to three dimensions.

\subsection{For 2d HOTRG}
\label{AM_two_dim}

\begin{figure}[t!]
    \centering
    \includegraphics[width=0.8\columnwidth, angle=0]{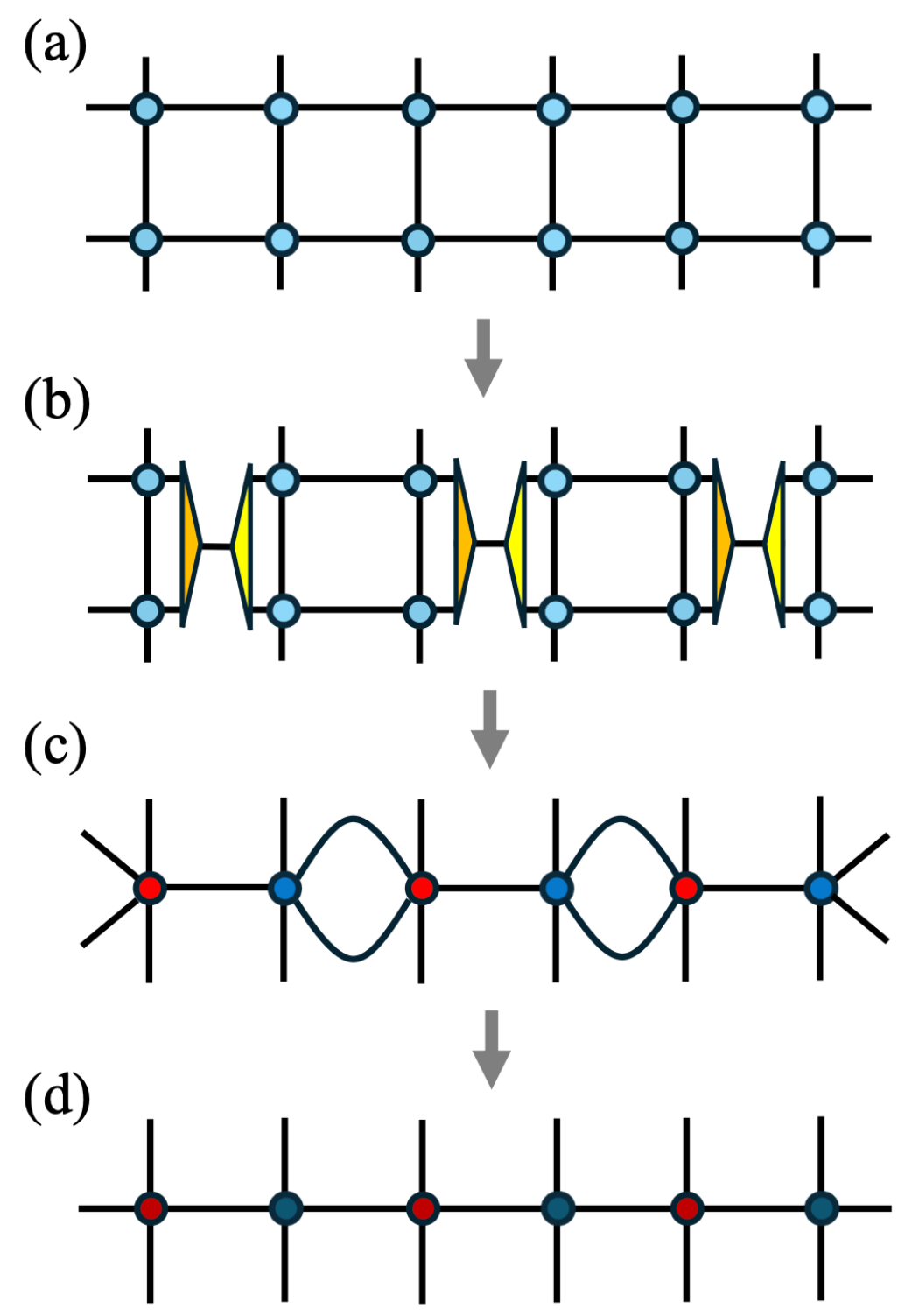}
    \caption{
        Schematic picture of the coarse-graining based on the all-mode boundary HOTRG in two-dimensional system.
        (a)$\rightarrow$(b): The squeezers are inserted in an alternative way following Eq.~(\ref{eqn:2d_2x2_squeezer}).
        (b)$\rightarrow$(c): Two original tensors $T$ and one of the squeezers are contracted to make new tensor $E$ or $O$ following Eq.~(\ref{eqn:2d_E_O}).
        (c)$\rightarrow$(d): Coarse-graining step from $E$ and $O$ to $E'$ and $O'$ in Eq.~(\ref{eq:second2d}) using additional stochastic noise.
        }
        \label{fig:2d_bHOTRG}
\end{figure}

In the subsection, we describe how to implement the all-mode technique using squeezers for the square lattice.
We begin with the homogeneous tensor network representation (see Fig.~\ref{fig:2d_bHOTRG}(a)) of the partition function $Z$ expressed in terms of local tensors, 
\be{
Z
=
\mathrm{Tr} \prod_{i} T_{x_ix_{i} 'y_iy_{i} '},
}
where $i$ runs all lattice sites, and the trace stands for the summation over all indices (bonds) $x_i,y_i$.
We assume that the bond dimension of the tensor is $D$.

We consider a fundamental $2\times 2$ lattice region which has a loop structure and insert a set of squeezers
\be{
\begin{aligned}
    \vcenter{\hbox{\includegraphics[width=7.5cm]{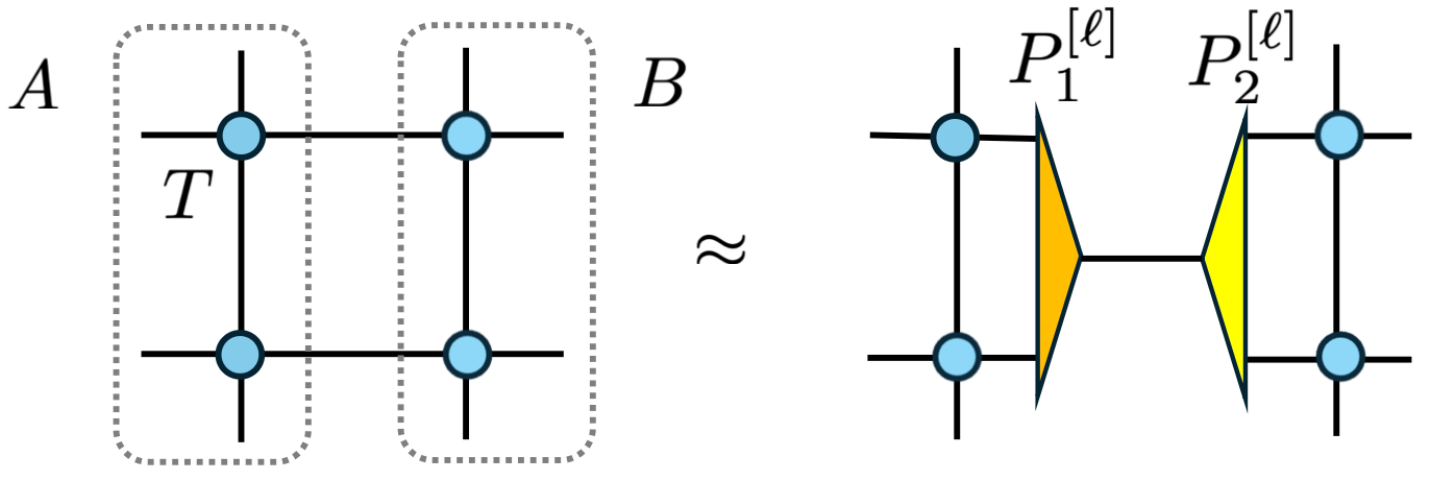}}}
\end{aligned}.
\label{eqn:2d_2x2_squeezer}
}
For a given noise variable, the squeezers $P_1^{[\ell]}$ and $P_2^{[\ell]}$ can be constructed following the procedure described in the previous section with a computational cost of $O(D^6)$.
We then insert these squeezers throughout the tensor network in an alternating pattern \footnote{We empirically find that this alternating pattern can control the error well.} as illustrated in Fig.~\ref{fig:2d_bHOTRG}(b).
Although the same colors are used in the figure for simplicity, an independent noise matrix $\xi$ is assigned to each pair of squeezers.
The new tensors $E$ and $O$ are then constructed by the contractions
\be{
\begin{aligned}
    \vcenter{\hbox{\includegraphics[width=5.cm]{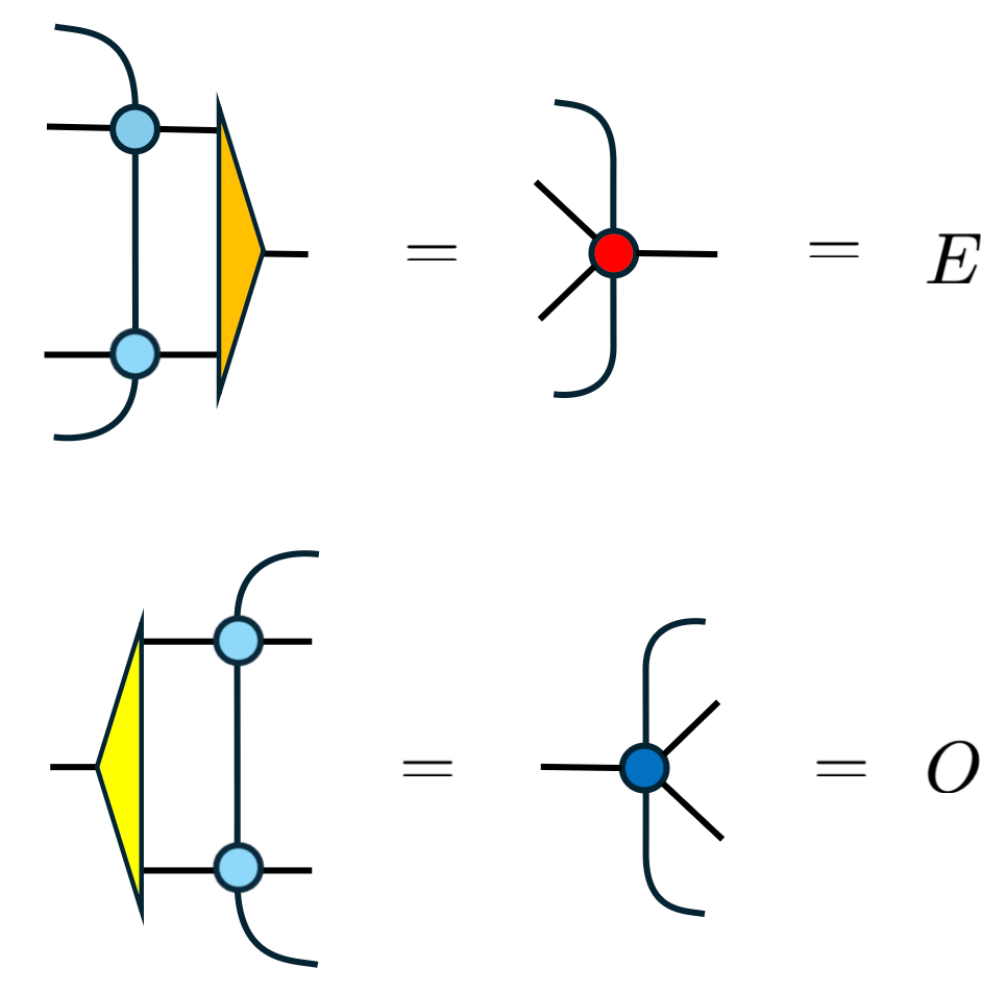}}}
\end{aligned},
\label{eqn:2d_E_O}
}
with computational cost of $O(D^7)$ per fundamental region.
Since this construction must be performed independently in each fundamental region of the tensor network, it is repeated $V/4$ times for a square lattice of volume $V$.
The resulting tensor network is shown in Fig.~\ref{fig:2d_bHOTRG}(c), where
the partition function is now given by the $E^{(i)}$ and $O^{(i)}$ for each new lattice site $i$,
\be{
Z
\approx
\mathrm{Tr}
\prod_{i} ^{V/4}
E^{(i)}
O^{(i)}.
}

The next task is to coarse-grain the other half of the indices. 
To this end, we introduce a new fundamental region composed of the $O$ and $E$,
\be{
\begin{aligned}
    \vcenter{\hbox{\includegraphics[width=4.cm]{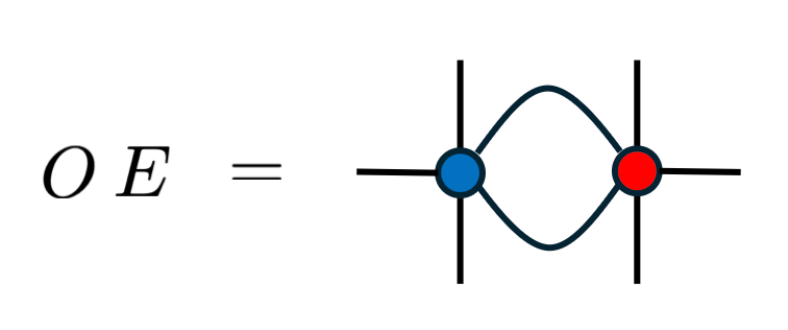}}}
\end{aligned}.
\label{eqn:2d_EO}
}
One may either insert an additional set of stochastic squeezers into the two inner bonds of the network shown above or, equivalently, apply the SVD repeatedly, following a procedure analogous to that used in the anisotropic TRG \cite{ATRG,boundaryHOTRG}. In both cases, the computational cost scales as $O(D^7)$ per fundamental region. In this work, we adopt the latter approach, which we describe below.

We first apply the SVD separately to $O$ and $E$, with a computational cost of $O(D^7)$,
\be{
\begin{aligned}
    \vcenter{\hbox{\includegraphics[width=5.5cm]{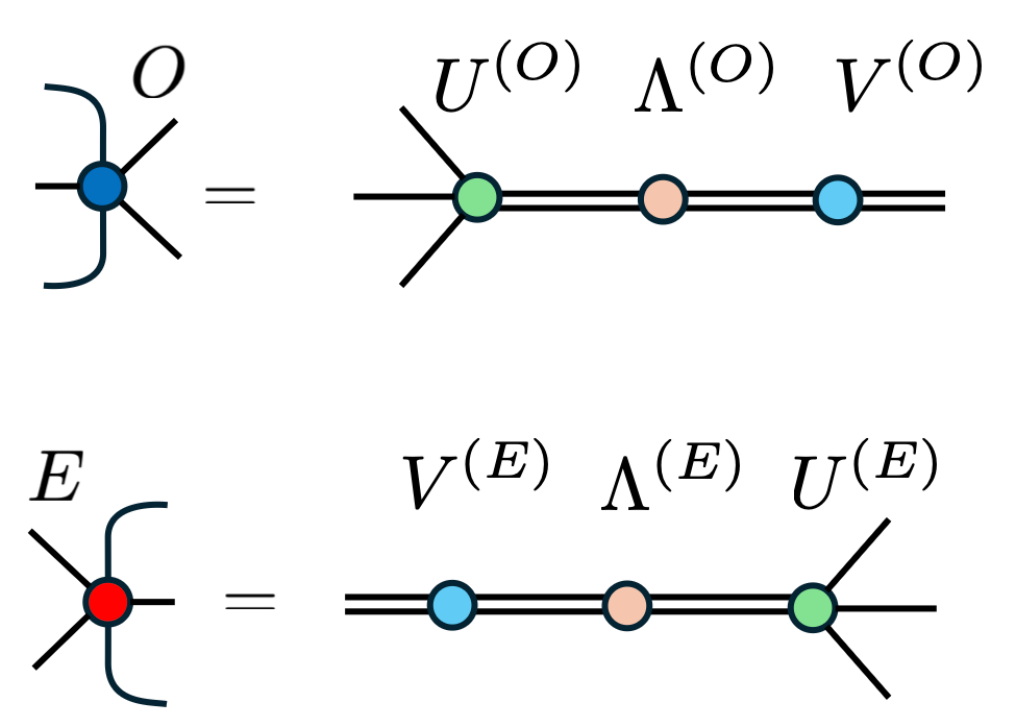}}}
\end{aligned}.
\label{eqn:2d_E_O_SVD}
}
Note that the indices of $O$ and $E$ are grouped differently from those in Eq.~\eqref{eqn:2d_E_O}. Their product can then be written as
\be{
OE=
U^{(O)}\Lambda^{(O)}V^{(O)}
V^{(E)}\Lambda^{(E)}U^{(E)}.
}
We then apply SVD to the middle part,
\be{
\Lambda^{(O)}
V^{(O)}
V^{(E)}
\Lambda^{(E)}
=
U^{(OE)}
\Lambda^{(OE)}
V^{(OE)}.
}
In this decomposition, we again introduce a new set of noise and the corresponding $\xi$.
When averaging over the noise ensemble, the following relation is expected to hold,
\beqnn{
\lefteqn{U^{(OE)}
\Lambda^{(OE)}
V^{(OE)}}\\
&=&
U^{(OE)}
\sqrt{\Lambda^{(OE)}}
\langle\xi\xi^\dagger\rangle
\sqrt{\Lambda^{(OE)}}
V^{(OE)}
+
O\left(\frac{1}{\sqrt{N_\mathrm{stat}}}\right).
\nonumber
}
In the end, for each $\ell$, we obtain the coarse-grained tensor $E'$ and $O'$,
\be{
\begin{aligned}
    \vcenter{\hbox{\includegraphics[width=7.cm]{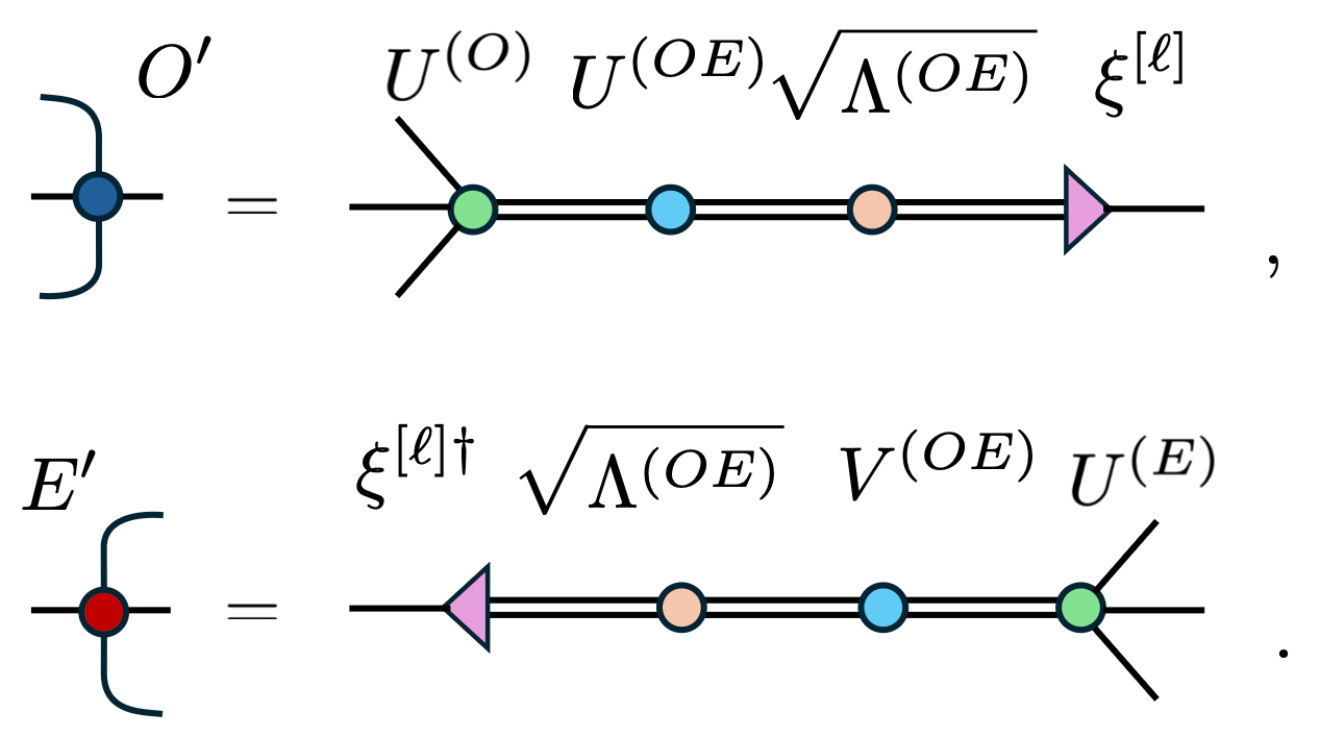}}}
\end{aligned}
\label{eq:second2d}
}
By repeating this procedure for all other fundamental regions,
the partition function is then expressed in terms of $E'$ and $O'$ as illustrated in Fig.~\ref{fig:2d_bHOTRG}(d),
\be{
Z
\approx
\mathrm{Tr}
\prod_i ^{V/4}
E' {}^{(i)}
O' {}^{(i)}.
}
In this coarse-graining procedure for one-direction where the number of tensors is reduced $V \to V/2$, the leading computational cost scales with $D^7\times(V/2)$.

The same procedure can be applied iteratively to carry out the coarse-graining. As in the conventional HOTRG, we first coarse-grain the tensor network along the $x$-direction and then along the $y$-direction. Summing over all coarse-graining steps, the number of local operations accumulates as
$V/2+V/4+V/8+\dots \approx V$,
so that the total computational cost is proportional to the lattice volume $V$. The coarse-graining procedure is repeated until the partition function is reduced to a $2\times2$ tensor network, whose final contraction requires only $O(D^6)$ computational cost.

Finally, we repeat the entire procedure using independent sets of noise variables and average the resulting estimates over $N_{\rm stat}$ samples,
\be{
Z\approx
\left\langle
\mathrm{Tr}
\prod_{i=1} ^{2}
E' {}^{(i)}
O' {}^{(i)}
\right\rangle.
\label{eqn:Z_2d_final}
}
The total computational cost scales as $O(D^7VN_{\rm stat})$. %where $V$ denotes the lattice volume.

\subsection{For 3d HOTRG}
\label{AM_three_dim}

\begin{figure}[tb!]
    \centering
    \includegraphics[width=0.95\columnwidth, angle=0]{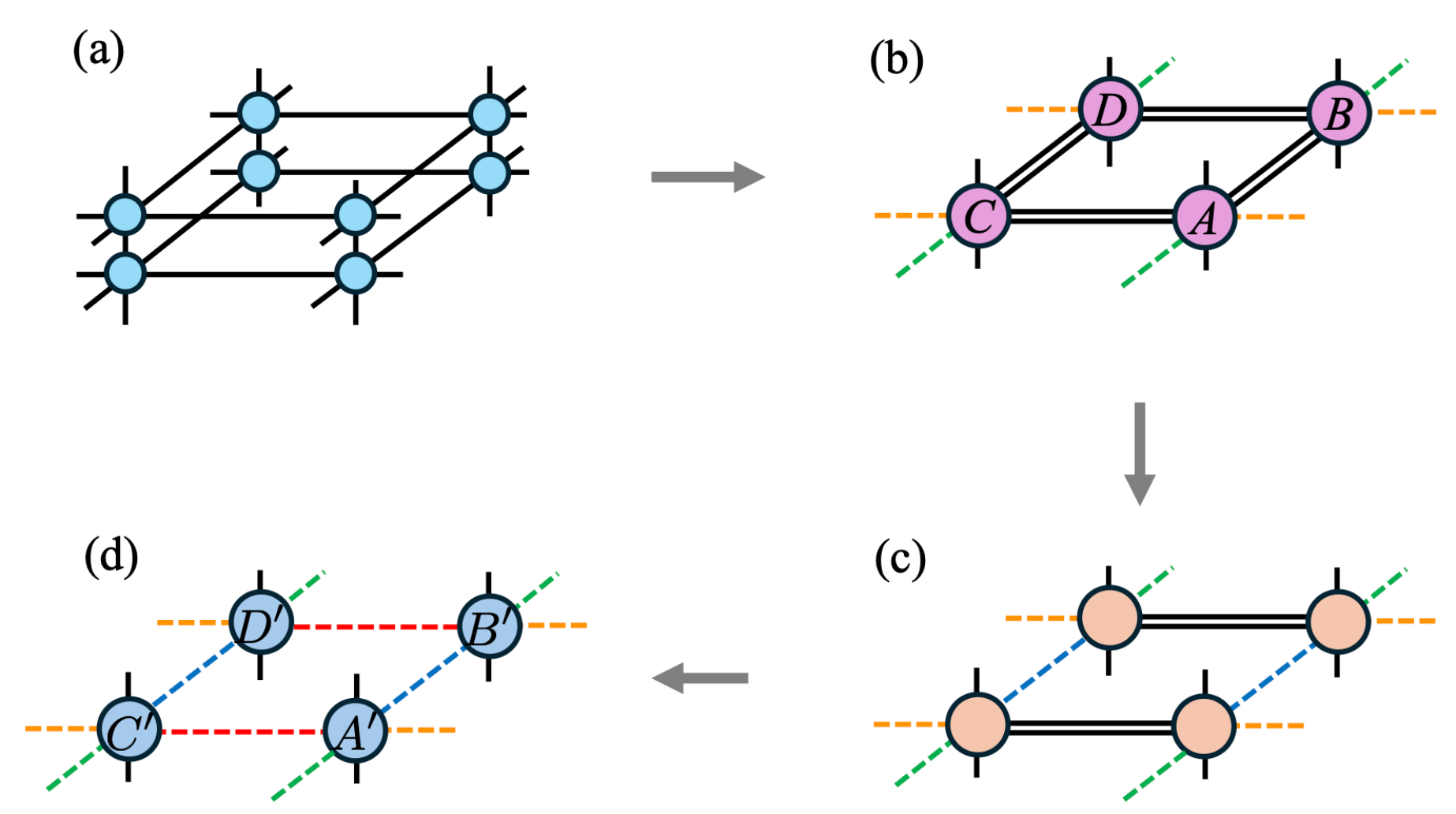}
    \caption{
        Schematic picture of the coarse-graining in three-dimensional system.
        (a)$\rightarrow$(b): Coarse-graining step %for even indices 
        from $T^{({\rm 3d})}$ to $A$, $B$, $C$, and $D$ using $P_{1,2}^{(x,y)}$ given in Eq.~(\ref{eq:first3d_x}) and (\ref{eq:first3d_y}).
        (b)$\rightarrow$(c): Coarse-graining for bonds between $A$ and $B$ ($C$ and $D$) using Eq.~(\ref{eq:3d_34x}).
        (c)$\rightarrow$(d): Coarse-graining for remaining bonds given in Eq.~(\ref{eq:3d_34y}) and producing $A'$, $B'$, $C'$, and $D'$.
        }
        \label{fig:3d_bHOTRG2}
\end{figure}

We extend our approach to cubic lattice systems, whose partition function is given by
\be{
Z
=
\mathrm{Tr}
\prod_{i = 1}
T_{x_ix_i'y_iy_i'z_iz_i'} ^\mathrm{(3d)},
}
where the bond dimension of the local tensor is assumed to be $D$.
We basically follow the standard three-dimensional HOTRG procedure. Namely, the tensor network is first coarse-grained along the $z$-direction, then along the $y$-direction, and so on.
We first consider a $2\times1\times2$ lattice as a fundamental region,
and construct a pair of stochastic squeezers, $P_1^{(x)}$ and $P_2^{(x)}$, on every other bond along the $x$-direction,
\be{
\begin{aligned}
    \vcenter{\hbox{\includegraphics[width=7.cm]{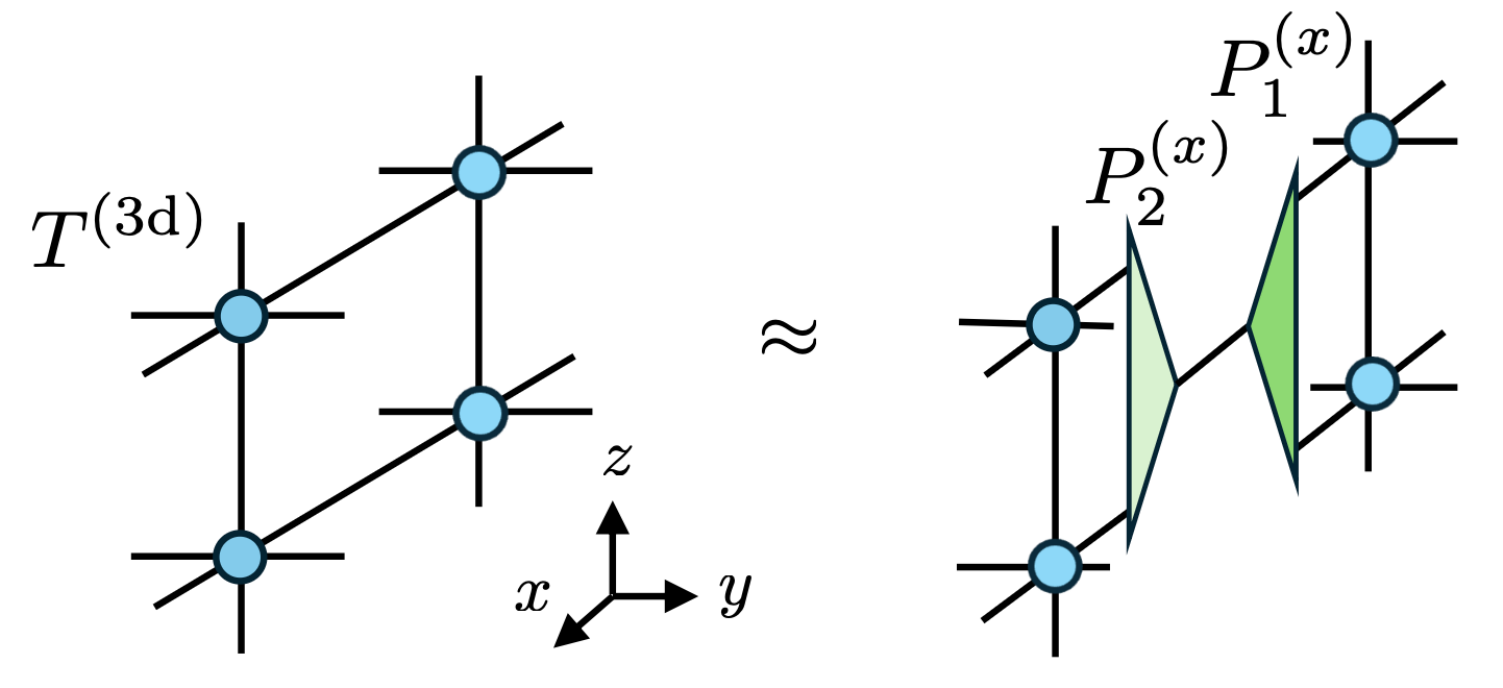}}}
\end{aligned}.
\label{eq:first3d_x}
}
In our implementation, we also insert similar pairs of squeezers at the corresponding positions translated along the $y$- and $z$-directions. As already noted in the two-dimensional case, the random variables associated with different pairs are generated independently. 
Nevertheless, in the following, corresponding squeezers are represented by the same notation and color.

As the next step, we consider a different fundamental region, namely a $1\times2\times2$
tensor network to which two $P_1^{(x)}$ squeezers determined in the previous step are attached,
and construct another pair of squeezers, $P_1^{(y)}$ and $P_2^{(y)}$, along the $y$-direction,
\be{
\begin{aligned}
    \vcenter{\hbox{\includegraphics[width=7.5cm]{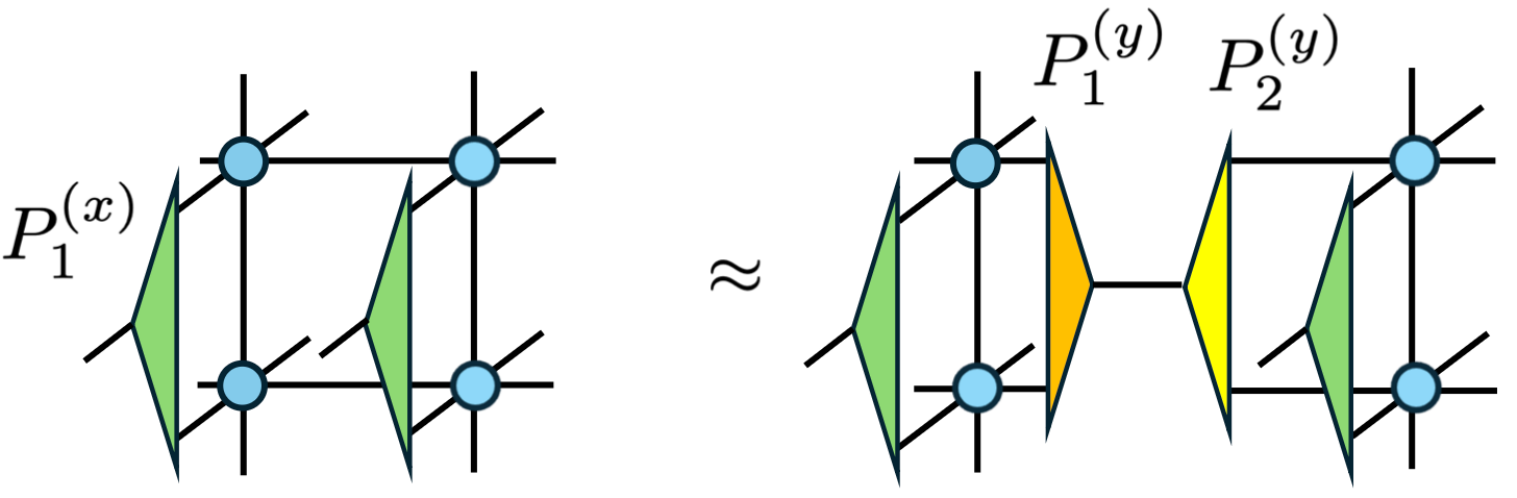}}}
\end{aligned}.
\label{eq:first3d_y}
}
In this way, $P_{1,2}^{(y)}$ are constructed using previously determined $P_1^{(x)}$. We use this nested construction since in two-dimensional case, it provides better control over the error.

After making the squeezers $P_1^{(x)}$ and $P_1^{(y)}$, we contract them with two
$T^{({\rm 3d})}$ tensors to obtain a partially coarse-grained tensor $A$ as follows,
\be{
\begin{aligned}
    \vcenter{\hbox{\includegraphics[width=6.cm]{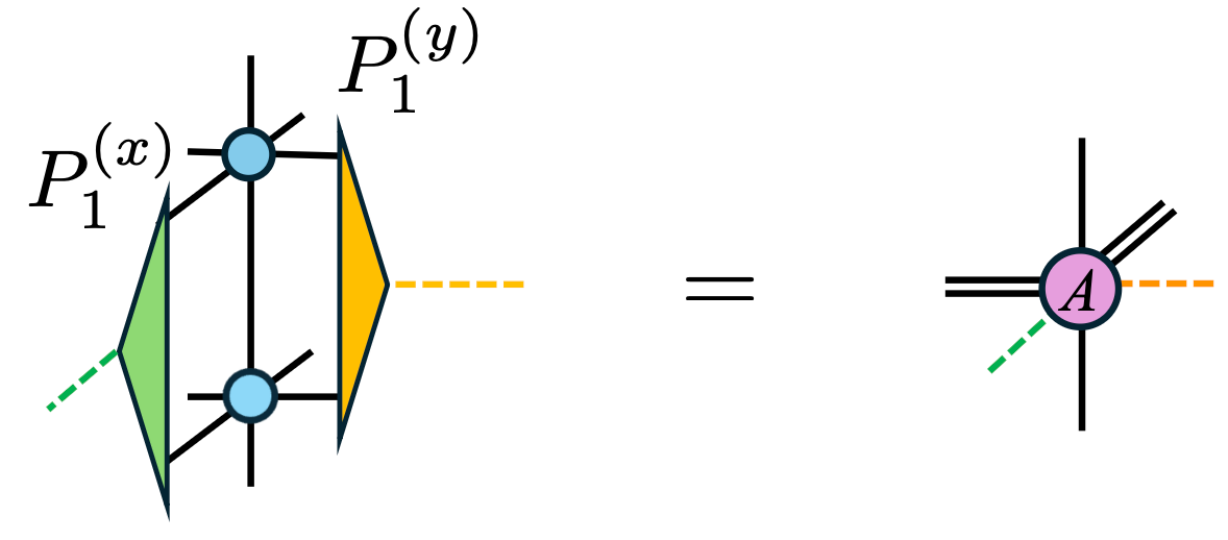}}}
\end{aligned}.
\label{eq:second3d}
}
This step has a computational cost of $O(D^{11})$, which constitutes the dominant part of this method.
We apply the same procedure to other corners of the $2\times2\times2$ lattice as shown in Fig.~\ref{fig:3d_bHOTRG2}(a)$\to$(b), to obtain partially coarse-grained tensors $B$, $C$, and $D$.
We emphasize again that the noise variables associated with different pairs of squeezers are generated independently, even though the same notation and color are used for the corresponding squeezers.

The remaining task is to coarse-grain the black double bonds shown in Fig.~\ref{fig:3d_bHOTRG2}(b).
To this end, we construct and insert an additional pair of stochastic squeezers, $P_3^{(x)}$ and $P_4^{(x)}$, between $A$ and $B$,
\be{
\begin{aligned}
    \vcenter{\hbox{\includegraphics[width=8cm]{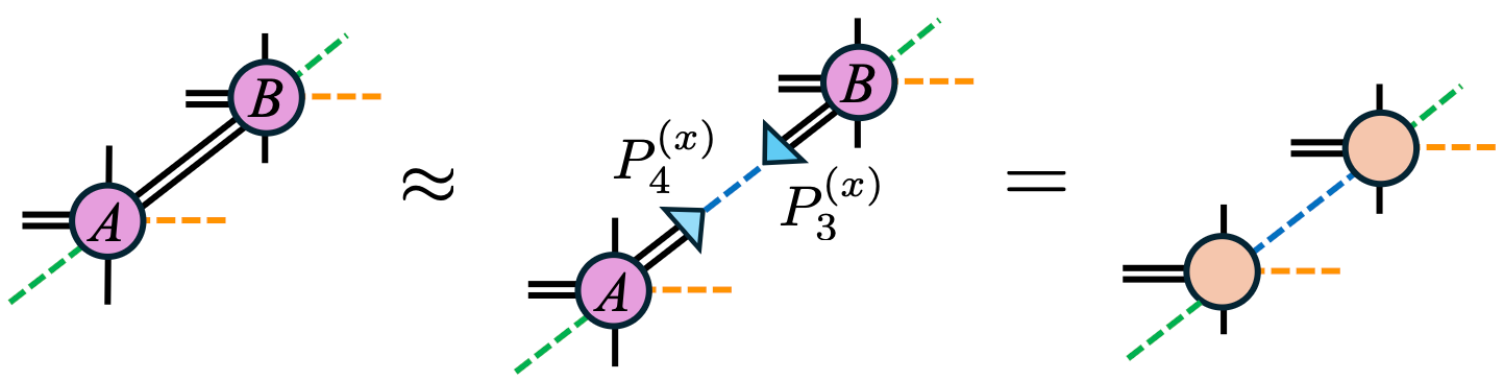}}}
\end{aligned}.
\label{eq:3d_34x}
}
We also insert a similar pair of squeezers between $C$ and $D$ using an independent set of noise variables.
Using the squeezers, the double bonds along the $x$-direction are coarse-grained as shown in the step from (b) to (c) in Fig.~\ref{fig:3d_bHOTRG2}.

As the final step, we insert another pair of squeezers on the remaining double bonds and obtain the coarse-grained tensor $A'$ and $C'$.
\be{
\begin{aligned}
    \vcenter{\hbox{\includegraphics[width=8.5cm]{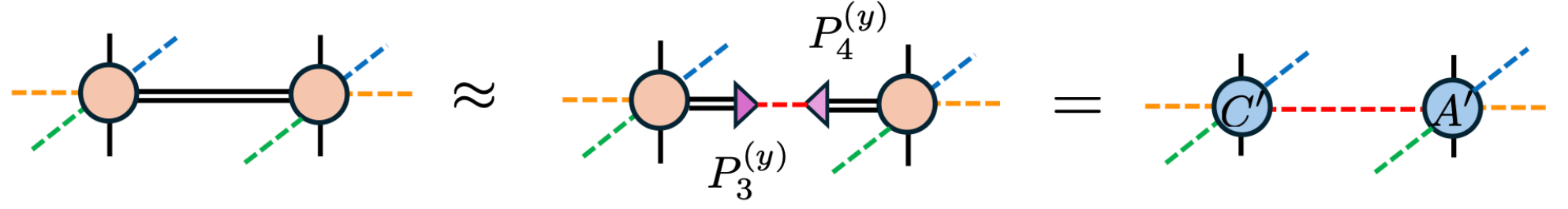}}}
\end{aligned}.
\label{eq:3d_34y}
}
A global view of this step is shown in Fig.~\ref{fig:3d_bHOTRG2}(c)$\to$(d), where the coarse-grained tensors $B'$ and $D'$ are also generated.
We repeat the same procedure for all positions translated along $y$- and $z$-directions.
The tensors $A'$, $B'$, $C'$ and $D'$ has same index structure index structure as $T$ and thus serve as the coarse-grained tensors representing the partition function,
\be{
Z
=
\mathrm{Tr}
\prod_{i} ^{V/8}
A^{'(i)}
B^{'(i)}
C^{'(i)}
D^{'(i)},
}
where $V$ denotes the volume of the initial cubic lattice.
This completes the coarse-graining along the $z$-direction, reducing the total number of tensors by a factor of two.

Repeating the coarse-graining procedure for $x$- and $y$-direction,
until the tensor network is reduced to $2\times2\times1$ whose contraction can be done with a computational cost of $O(D^6)$ and this is negligible in the overall computational cost.
The entire procedure is repeated for $N_{\rm stat}$ independent noise realizations, and the partition function is estimated as
\be{
Z
\approx
\left\langle
\mathrm{Tr}
A^{'}
B^{'}
C^{'}
D^{'}
\right\rangle.
}
The total computational cost scales as $O(D^{11}VN_{\rm stat})$.

\section{Numerical results}
\label{result}

\subsection{2d Ising model}

In the subsection, we demonstrate our method using the Ising model on the square lattice. Its partition function is represented as
\be{
Z
=
\mathrm{Tr}
\prod_{i = 1}
T_{x_ix_i'y_iy_i'} ^\mathrm{(Ising)}
}
where the initial tensor $T^\mathrm{(Ising)}$, with bond dimension two, is defined by
\begin{align}
T_{x_ix_i'y_iy_i'} ^\mathrm{(Ising)}
=
\sum_{\alpha=1}^2
	W_{\alpha,x_i}
	W_{\alpha,x_i'}
	W_{\alpha,y_i}
	W_{\alpha,y_i'},
\end{align}
with
\be{
	W_{\alpha,x}
		=
		\begin{pmatrix}
    		 \sqrt{\mathrm{cosh}\beta} & {\phantom{-}\sqrt{\mathrm{sinh}\beta}} \\
    		 {\sqrt{\mathrm{cosh}\beta}} & {-\sqrt{\mathrm{sinh}\beta}}
        \end{pmatrix},
\label{eqn:W}
}
and $\beta$ denotes the inverse temperature.
The critical inverse temperature is exactly known to be $\beta_{\rm c} = \frac{\mathrm{log}(1+\sqrt{2})}{2}$.

As described in the previous section, for each realization of the noise variables,
we coarse-grain the initial tensor network until it is reduced to $2\times2$ tensor network and then perform the final contraction.
The resulting estimates are averaged over $N_{\rm stat}$ independent noise realizations as in Eq.~\eqref{eqn:Z_2d_final}.
The statistical errors are estimated using the jackknife method as described in Appendix \ref{sec:JK}.

\begin{figure}[htbp]
\begin{center}
 \includegraphics[width=0.95\columnwidth, angle=0]{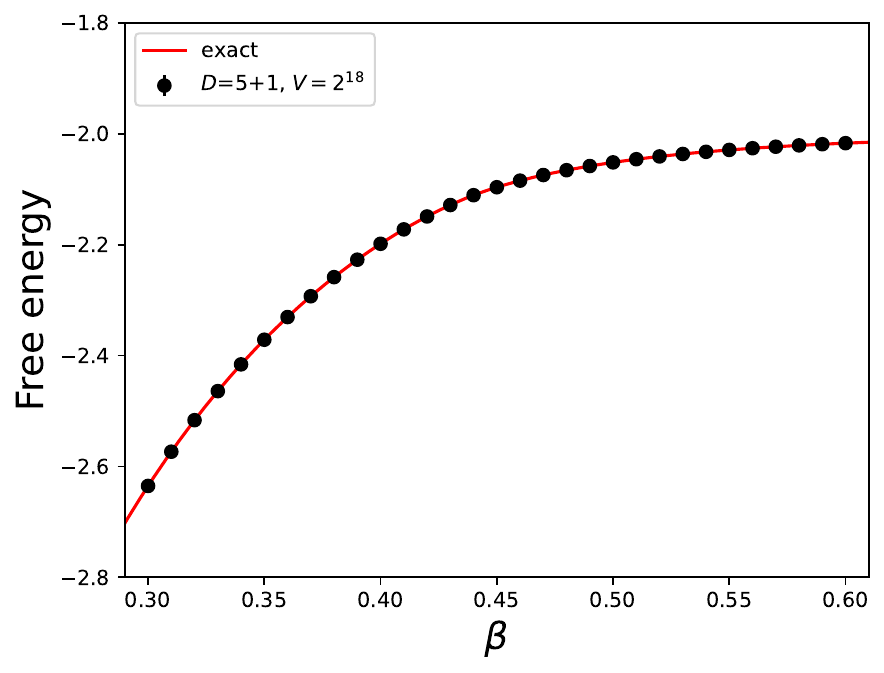}
 \caption{
 Free energy density $f$ of the two-dimensional Ising model obtained using all-mode HOTRG for the lattice volume $V = 2^{9}\times2^9$, the bond dimension $D=5+1$, and $N_{\rm stat}=100$. The red curve represents the exact solution in the thermodynamic limit.
}
\label{fig:2dIsing_free}
\end{center}
\end{figure}

\begin{figure}[htbp]
\begin{center}
 \includegraphics[width=0.95\columnwidth, angle=0]{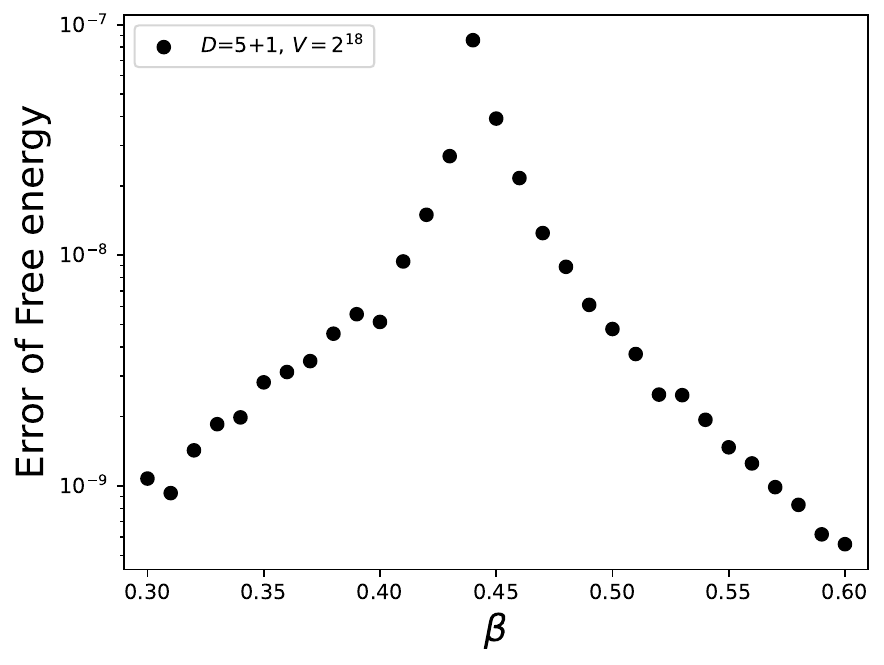}
 \caption{
 Statistical error in the free energy density obtained using all-mode HOTRG for the volume $V = 2^{9}\times 2^9$, the bond dimension $D=5+1$, and $N_{\rm stat}=100$.
}
\label{fig:2dIsing_free_err}
\end{center}
\end{figure}

We first present the free energy density
\be{
f = -\frac{1}{V\beta}\mathrm{log}Z
}
as a function of $\beta$ in Fig.~\ref{fig:2dIsing_free} where the all-mode result is compared with the exact solution.
The lattice volume, the bond dimension, and the total number of samples are set to $V=2^9\times 2^9$, $D=5+1$, and $N_{\rm stat}=100$, respectively.
The statistical error bars are too small to be seen on this scale of this figure.
No visible difference can be observed between the two results at this scale. 
Figure \ref{fig:2dIsing_free_err} shows the statistical error at each data point.
The error increases near the critical point, exhibiting a trend similar to that of the systematic error in the conventional TRG and HOTRG methods.
Since the statistical error can be evaluated directly within our method, it may provide a useful indicator for roughly locating the critical point.

\begin{figure}[htbp]
\begin{center}
 \includegraphics[width=8cm, angle=0]{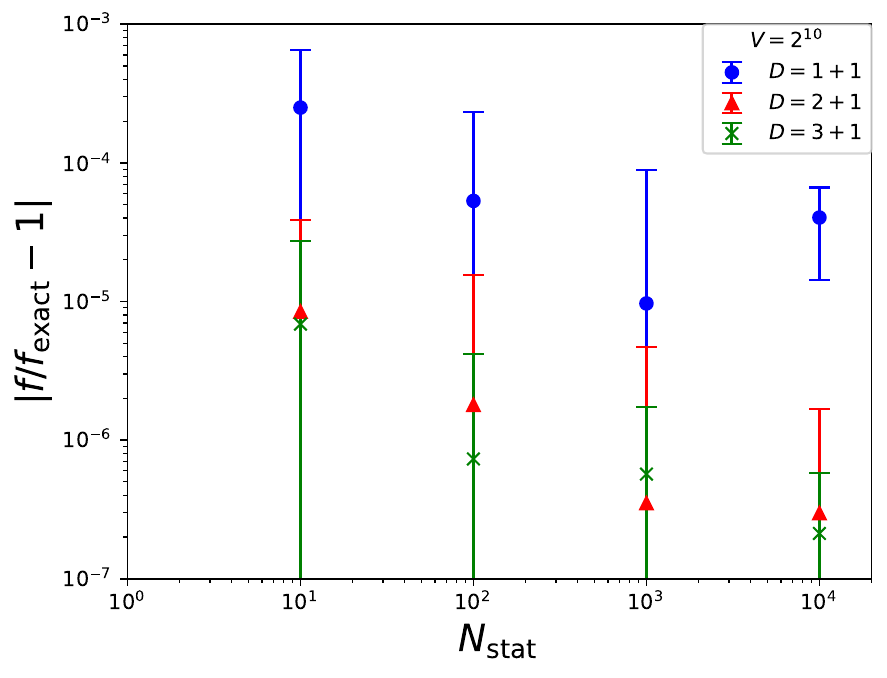}
 \caption{Relative deviation of the free energy in Eq.~\eqref{eqn:relative_deviation} as a function of $N_{\rm stat}$ for $V = 2^{5}\times2^5$ at the critical point.
}
\label{fig:V_P}
\end{center}
\end{figure}

\begin{figure}[htbp]
\begin{center}
 \includegraphics[width=8cm, angle=0]{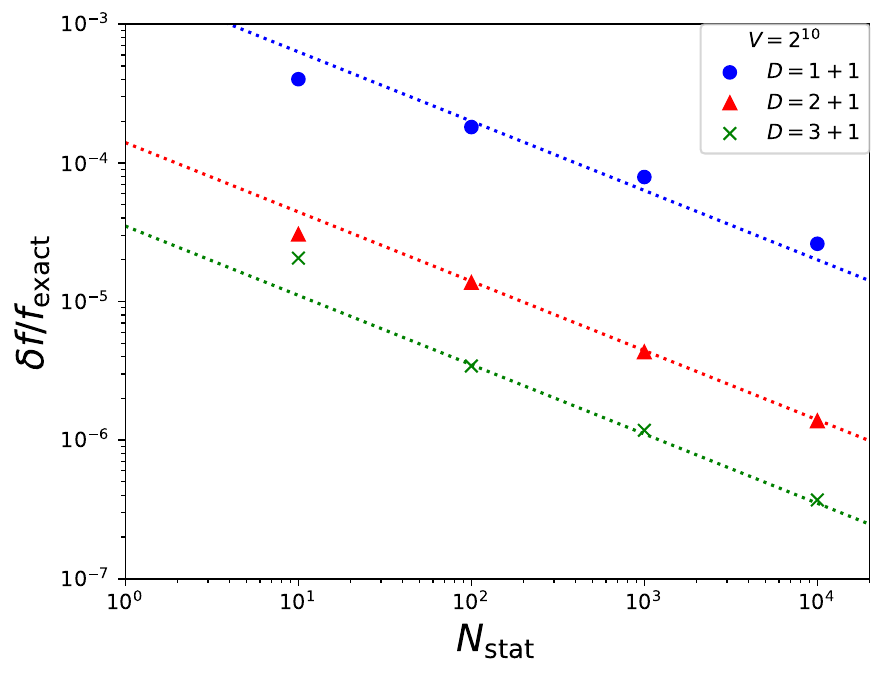}
 \caption{Statistical error in the free energy normalized by its exact value. All parameters are the same as those in Fig.~\ref{fig:V_P}.
 The dotted lines are proportional to $1/\sqrt{N_\mathrm{stat}}$.
}
\label{fig:V_P_err}
\end{center}
\end{figure}

\begin{figure}[htbp]
\begin{center}
 \includegraphics[width=8cm, angle=0]{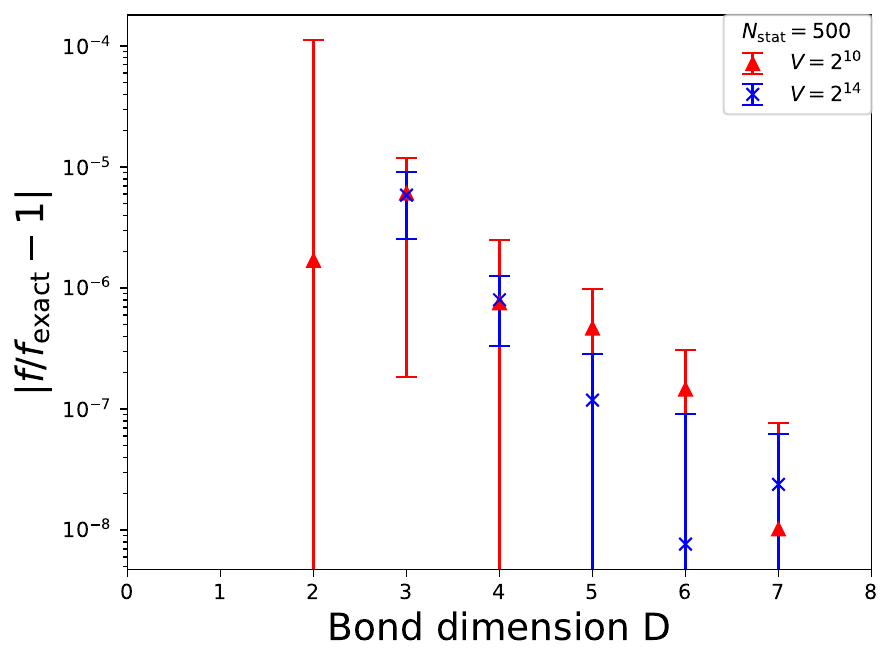}
 \caption{Relative deviation of the free energy as a function of $D$ for $N_\mathrm{stat} = 500$ at the critical point. The lattice volume is varied $V = 2^{5}\times 2^5$ (triangle) and $V=2^{7}\times 2^7$ (cross).
}
\label{fig:D_P}
\end{center}
\end{figure}

To assess the validity of our statistical method more carefully, we plot the relative deviation of the free energy,
\be{
\frac{f}{f_{\rm exact}}-1,
\label{eqn:relative_deviation}
}
as a function of $N_{\rm stat}$ in Fig.~\ref{fig:V_P} for several bond dimensions $D=2,3,4$.
The lattice volume and inverse temperature are fixed at $V=2^5\times 2^5$ and the critical value $\beta_{\rm c}$, respectively.
In most cases, the relative deviation decreases with increasing $N_{\rm stat}$ and becomes consistent with zero within the statistical errors. Note that the smallest bond dimension, $D=1+1$, may be too small to reach convergence.
We next examine the scaling of the statistical errors with $N_{\rm stat}$ in more detail.
Figure \ref{fig:V_P_err} shows the expected convergence behavior, namely, the statistical error scales as $1/\sqrt{N_\mathrm{stat}}$.
All parameters are the same as those used in Fig.~\ref{fig:V_P}.
We then investigate the dependence of the relative deviation on the bond dimension, as shown in Fig.~\ref{fig:D_P}.
The number of samples and inverse temperature are fixed to $N_\mathrm{stat} = 500$, and $\beta=\beta_{\rm c}$, respectively, while
the lattice volume is varied between $V=2^5\times 2^5$ and $2^7\times2^7$. 
As seen in the figure, the accuracy improves exponentially with increasing $D$ in the large $D$-regime.
A similar trend is observed even for the larger volume.

\begin{figure}[htbp]
\begin{center}
 \includegraphics[width=8cm, angle=0]{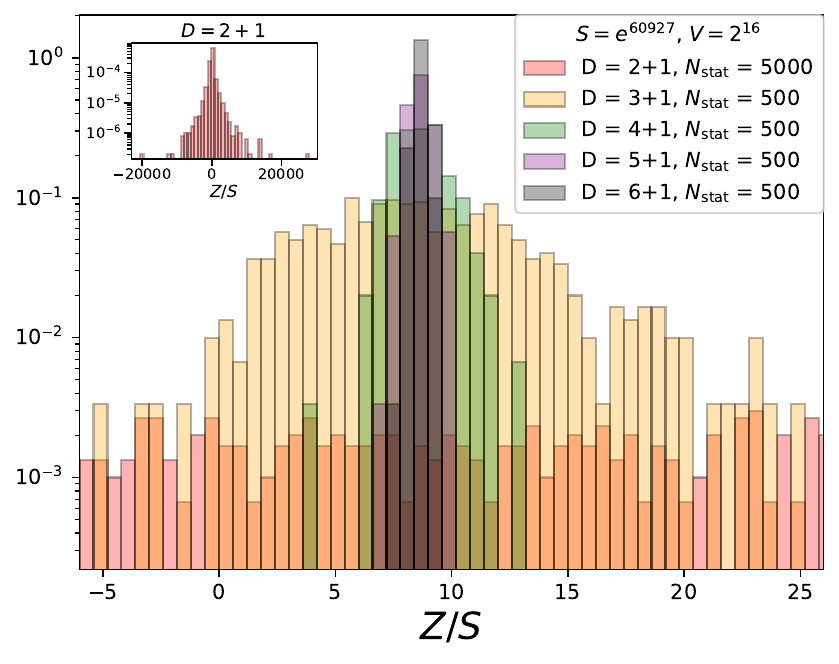}
\caption{Normalized histogram of the partition function $Z$ with a rescaling factor $S = e^{60927}$ in the volume $V = 2^{8}\times 2^8$ with bond dimensions $D = 3,4,5,6,7$.
}
\label{fig:hist_all}
\end{center}
\end{figure}

The convergence discussed above is subject to a practical limitation associated with the distribution of the stochastic estimator of the partition function, as discussed in Ref.~\cite{Allmode}. For small bond dimensions and insufficient statistics, the distribution of $Z$ becomes broad and contains a significant fraction of negative samples. As a result, the sample average or some leave-one-out jackknife estimates of $Z$ may become nonpositive, rendering the real-valued free energy density ill defined (see Appendix~\ref{sec:JK}). This explains why a much larger number of samples is required to achieve convergence at the smallest bond dimension.
This problem is substantially alleviated by increasing the bond dimension. As shown in Fig.~\ref{fig:hist_all}, the fluctuations of $Z$ are significantly suppressed as the bond dimension increases. On the other hand, since the number of stochastic variables increases with the lattice volume, this issue is expected to become more severe for larger systems.
We note that all data sets presented in this paper satisfy the positivity condition in Eq.~\eqref{eq:JK_positivity}, except for the small-bond-dimension data in Fig.~\ref{fig:hist_all}, which were obtained at the relatively large volume $V=2^{16}$.

\begin{figure}[htbp]
\begin{center}
  \includegraphics[width=0.95\columnwidth, angle=0]{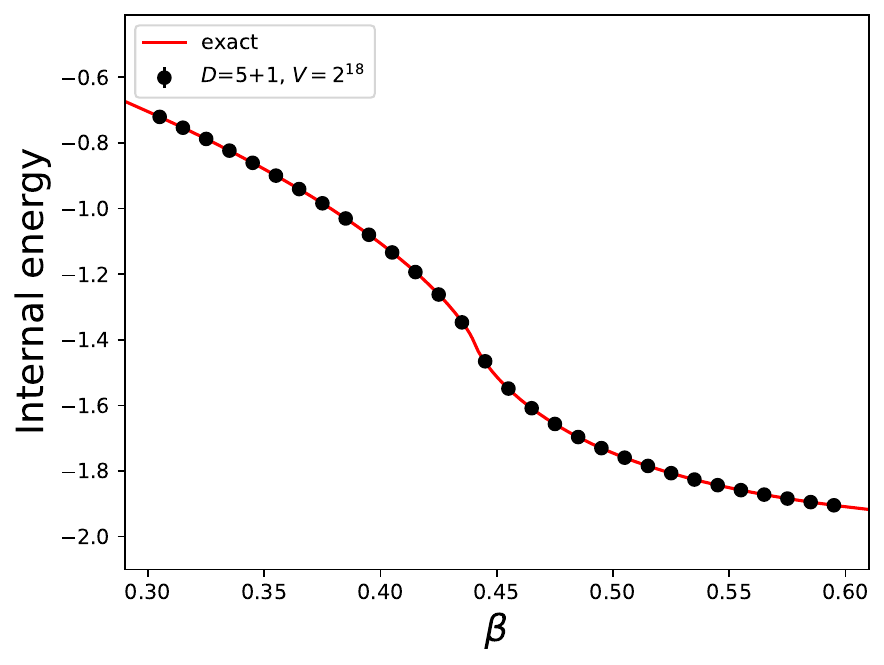}
 \caption{
 Internal energy density obtained using all-mode HOTRG for the volume $V = 2^{9}\times 2^9$, the bond dimension $D=5+1$, and $N_{\rm stat}=100$.
 The red curve represents the exact solution.
}
\label{fig:2dIsing_inner}
\end{center}
\end{figure}

\begin{figure}[htbp]
\begin{center}
 \includegraphics[width=0.95\columnwidth, angle=0]{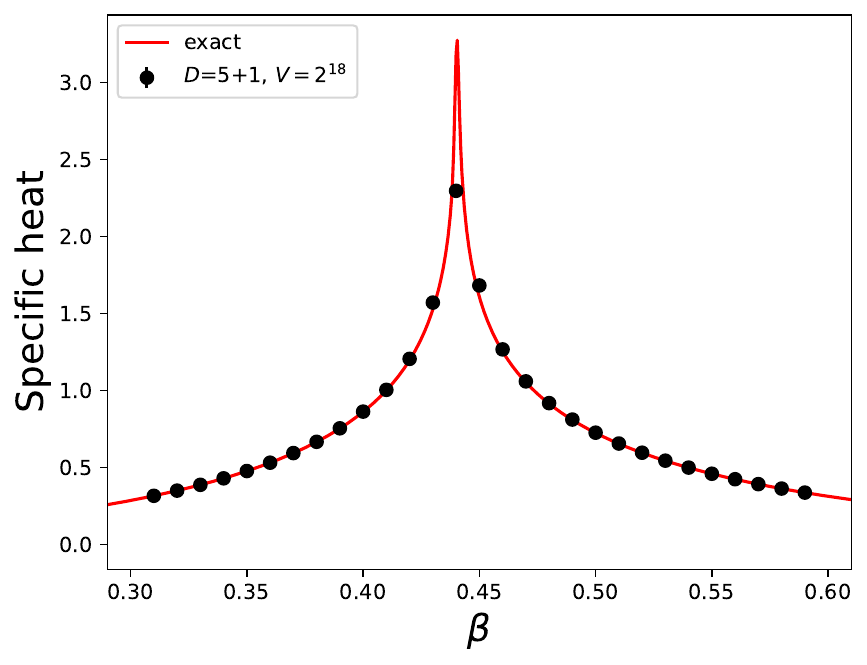} 
 \caption{
 Specific heat obtained using all-mode HOTRG with the same parameters as those in Fig.~\ref{fig:2dIsing_inner}.
}
\label{fig:2dIsing_spec}
\end{center}
\end{figure}

We next compute the internal energy density
\be{
u = -\frac{1}{V}\frac{\partial \mathrm{log}Z}{\partial \beta}
} and the specific heat per site,
\be{
c= -\beta^2\frac{\partial u}{\partial \beta}
}
and compare them with the exact results.
As shown in \ref{fig:2dIsing_inner} and \ref{fig:2dIsing_spec}, our results remain in good agreement with the exact solutions for both the internal energy and the specific heat.
Here we use the numerical derivative to estimate the first and second derivatives.
Note that we do not employ the impurity-tensor method to evaluate these derivatives, since it does not improve the accuracy compared with simple numerical differentiation.

\subsection{3d Ising model}

We also demonstrate our method using the three-dimensional Ising model.
Its partition function is given in terms of the tensor network representation,
\be{
Z
=
\mathrm{Tr}
\prod_{i = 1}
T_{x_ix_i'y_iy_i'z_iz_i'} ^\mathrm{(3d)}
}
where the initial tensor $T^\mathrm{(3d)}$ is constructed as
\begin{align}
T_{x_ix_i'y_iy_i'z_iz_i'} ^\mathrm{(3d)}
=	\sum_{\alpha=1}^2
	W_{\alpha,x_i}
	W_{\alpha,x_i'}
	W_{\alpha,y_i}
	W_{\alpha,y_i'}
	W_{\alpha,z_i}
	W_{\alpha,z_i'},
\end{align}
and the $2\times2$ matrix $W$ is given in Eq.~\eqref{eqn:W}.

\begin{figure}[htbp]
\begin{center}

\includegraphics[width=0.95\columnwidth, angle=0]{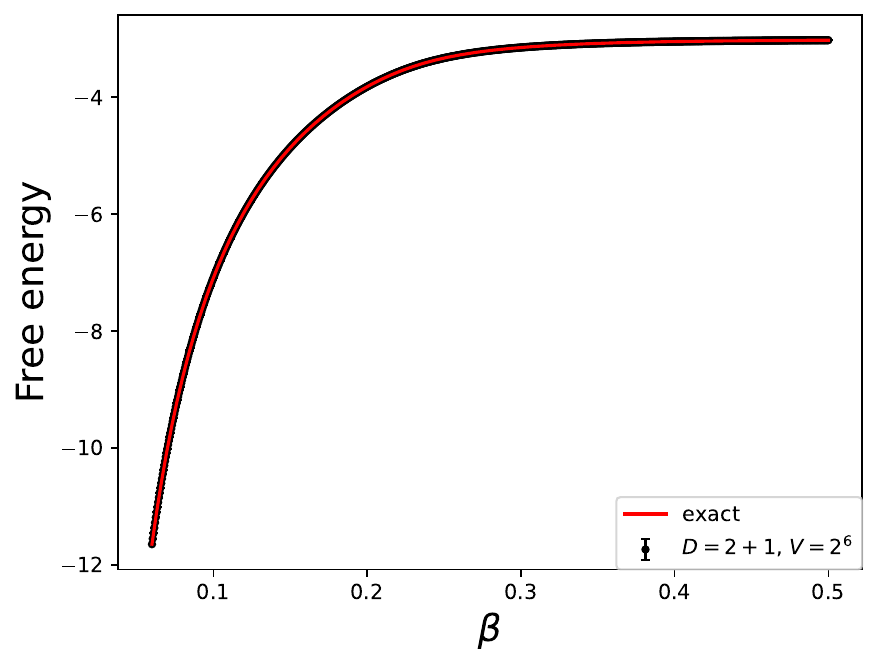}
 \caption{
Free energy density obtained using all-mode HOTRG for the volume $V = 4\times 4 \times 4$, the bond dimension $D=2+1$, and $N_{\rm stat}=100$.
 The red curve represents the exact solution.
}
\label{fig:3dIsing_free}
\end{center}
\end{figure}

\begin{figure}[htbp]
\begin{center}

\includegraphics[width=0.95\columnwidth, angle=0]{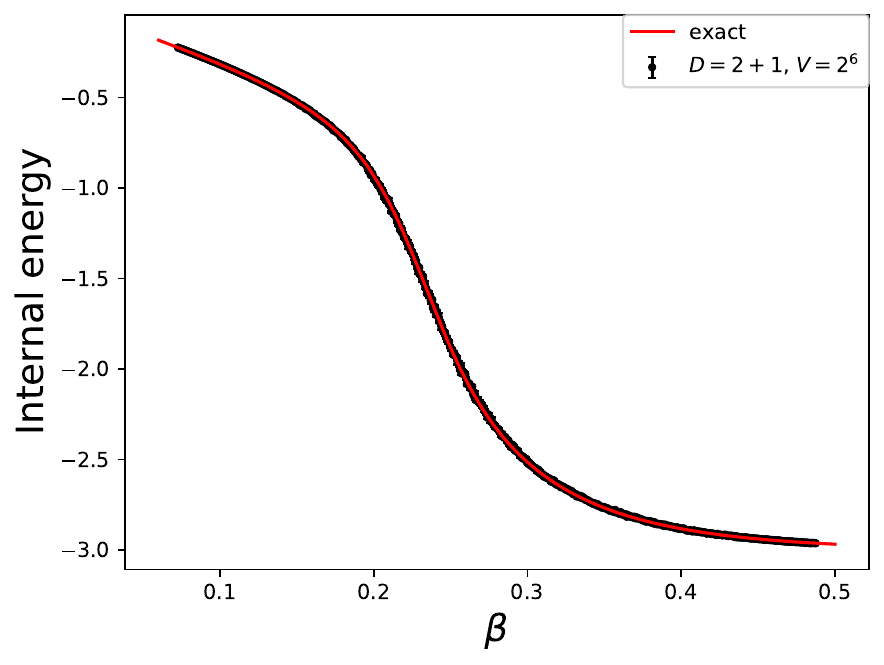}
 \caption{
Internal energy density obtained using all-mode HOTRG.
The parameters are the same as Fig.~\ref{fig:3dIsing_free}.
}
\label{fig:3dIsing_inner}
\end{center}
\end{figure}

\begin{figure}[htbp]
\begin{center}

\includegraphics[width=0.95\columnwidth, angle=0]{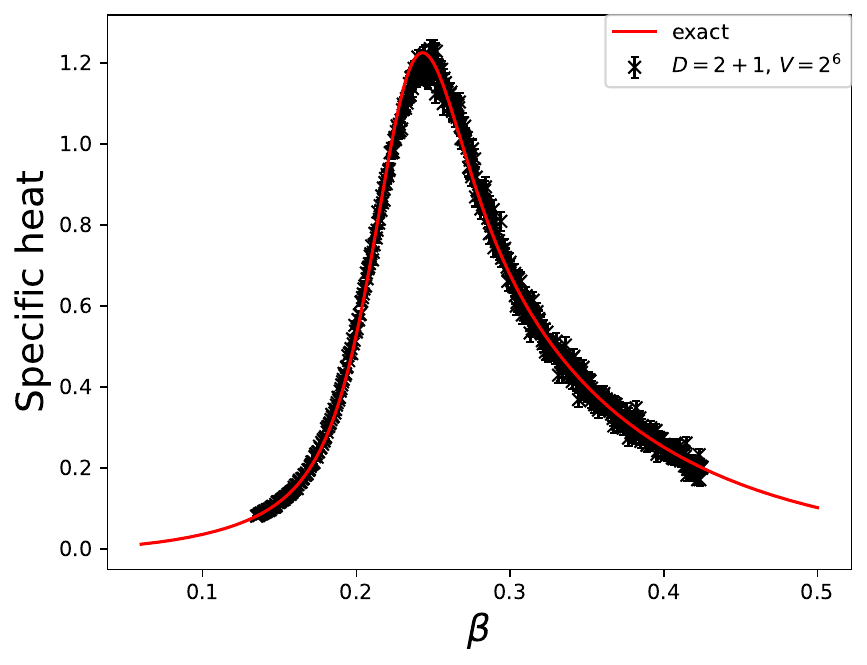}
 \caption{
Specific heat obtained using all-mode HOTRG.
The parameters are the same as Fig.~\ref{fig:3dIsing_free}.
}
\label{fig:3dIsing_spec}
\end{center}
\end{figure}

Since the three-dimensional Ising model is not exactly solvable in general, we focus here on a small system with volume $V=4\times4\times4$, for which exact results are available~\cite{PhysRevB.26.6285}.
For this volume, our results for the free energy density, internal energy density, and specific heat are shown in Figs.~\ref{fig:3dIsing_free}, \ref{fig:3dIsing_inner}, and \ref{fig:3dIsing_spec}, respectively.
The bond dimension is set to $D=2+1$, and the number of statistical samples is $N_{\rm stat}=100$.
Over a wide range of $\beta$, our results are consistent with the exact values within statistical errors.

For a more quantitative comparison, we consider the free energy density near the critical temperature, $T_{\rm c}^{(3{\rm d})}=1/\beta_{\rm c}^{(3{\rm d})}=4.5115$.
For the same parameter set, $V=4\times4\times4$, $N_{\rm stat}=100$, and $D=2+1$, the all-mode method yields
$f^{(3{\rm d})}=-3.55456(58)$,
which is in good agreement with the exact result,
$f_{\rm exact}^{(3{\rm d})}=-3.554768547$.

%{\color{red} we want more. relative deviation for 4x2x2, 4x4x2, 4x4x4 in slide of JPS meeting 2025!
%Furthermore, comparison of $f,u,c$ with exact solution on 4x4x4 in the same slides.}

%\begin{figure}[htbp]
%\begin{center}
%\includegraphics[width=0.95\columnwidth, angle=0]{./3d_plots/PLOT_V456_2_stat10000.pdf}
% \caption{
%elative deviation of the free energy as a function of the step in $V = 2^\mathrm{step}$, the bond dimension $D=2+1$, and $N_{\rm stat}=10000$.
% The dotted line represents the exact solution.
%}
%\label{fig:2dIsing_error}
%\end{center}
%\end{figure}

\section{Conclusion}
\label{concl}

In this work, we have generalized the stochastic all-mode technique to tensor-network coarse-graining schemes formulated in terms of squeezers. The central idea is to replace the deterministic truncation used in constructing the squeezers with a stochastic decomposition that retains contributions from all singular modes. The dominant modes are treated exactly, while the remaining modes are incorporated stochastically using random noise. In the limit of an infinite number of noise samples, the exact tensor-network contraction is recovered. In this way, the systematic error arising from the bond-dimension truncation is replaced by a statistical error that can be directly estimated from an ensemble of independent noise realizations.

We implemented this framework within the boundary-HOTRG formulation in both two and three dimensions. The total computational cost scales as $O(VN_{\rm stat}D^7)$ in two dimensions and as $O(VN_{\rm stat}D^{11})$ in three dimensions. Numerical calculations for the two-dimensional Ising model showed that the free energy density obtained with the all-mode HOTRG method agrees with the exact solution within statistical uncertainties. 
We also confirmed the expected $1/\sqrt{N_{\rm stat}}$ scaling of the statistical error and found that, at fixed $N_{\rm stat}$, the statistical error exponentially decreases with increasing bond dimension.
Furthermore, the internal energy density and specific heat, evaluated by numerical differentiation of the free energy, were found to be consistent with their exact values. As an initial test in three dimensions, we applied the method to the Ising model on a small $4^3$ lattice and found that the free energy, internal energy density, and specific heat are all in good agreement with the exact results.

An important practical limitation of the present approach is the 
broad distribution of the stochastic estimator of the partition 
function. For small bond dimensions or large lattice volumes, 
the fluctuations of the partition function can become large, and 
individual samples, or even leave-one-out jackknife averages, may 
take nonpositive values. In such cases, the logarithm required for 
the free energy density is not well defined, and a reliable error 
estimate requires a substantially larger number of samples. Our 
results indicate that increasing the bond dimension suppresses 
these fluctuations, but the required computational cost becomes 
particularly demanding in three dimensions.

The present formulation is not restricted to HOTRG. Since the 
all-mode construction is expressed directly in terms of 
squeezers, it can in principle be incorporated into other 
tensor-renormalization algorithms based on similar compression 
structures, including the anisotropic TRG, the minimally 
decomposed TRG, and the triad TRG. Applying the method to these 
computationally less expensive formulations may make it possible 
to employ larger bond dimensions and thereby control the broad 
distribution more effectively. Developing further techniques to 
reduce the variance of the stochastic estimator without 
reintroducing truncation bias will be essential for extending the 
all-mode approach to larger three-dimensional systems and to more 
complicated lattice models.

\section*{Acknowledgments}
We would like to thank H.~Ohki and M.~Tomii for helpful and insightful discussions.
This work was supported by JSPS KAKENHI Grant Number 21K03531, %ST
22H05251, %ST
24K17059, %KN
24H00214, %ST
and 25K07280. %ST

\appendix

\section{Jackknife analysis of the free energy density}
\label{sec:JK}

In the all-mode method, an estimate of the partition
function depends on the random noise introduced in each statistical
sample. Let $Z_i$, with $i=1,2,\ldots,N_{\rm stat}$, denote the estimate of
the partition function obtained from the $i$th sample. 
The estimator constructed from all samples is given by
\begin{equation}
\overline{Z}
=
\frac{1}{N_{\rm stat}}
\sum_{i=1}^{N_{\rm stat}} Z_i ,
\label{eq:JK_mean_Z}
\end{equation}
and the corresponding estimator of the free energy density is given by
\begin{equation}
\widehat{f}
=f(\bar Z)=
-\frac{1}{\beta V}\log \overline{Z}.
\label{eq:JK_full_f}
\end{equation}

Although
the exact partition function is strictly positive, an individual
estimate $Z_i$ may occasionally be negative because of statistical
fluctuations. Consequently, the free energy density cannot be evaluated
separately for each sample through
\begin{equation}
f_i=f(Z_i)=-\frac{1}{\beta V}\log Z_i
\end{equation}
when $Z_i\leq 0$.
To avoid taking the logarithm of each individual estimate, we apply the
jackknife method to the partition function before evaluating the
free energy density. 
For each $k=1,2,\ldots,N_{\rm stat}$, the leave-one-out jackknife
estimate of the partition function is defined by
\begin{equation}
Z^{(k)}
=
\frac{1}{N_{\rm stat}-1}
\sum_{\substack{i\neq k}}^{N_{\rm stat}} Z_i .
\label{eq:JK_leave_one_out_Z}
\end{equation}
The corresponding jackknife replicate of the free energy density is given by
\begin{equation}
f^{(k)}
=
f(Z^{(k)})=
-\frac{1}{\beta V}\log Z^{(k)}.
\label{eq:JK_leave_one_out_f}
\end{equation}
Since each $Z^{(k)}$ is an average over $N_{\rm stat}-1$ samples,
its fluctuations are smaller than those of the individual $Z_i$.
Thus, all $Z^{(k)}$ may remain positive even when some of the
individual estimates $Z_i$ are negative.

The average of the jackknife replicates is
\begin{equation}
\overline{f}_{\rm JK}
=
\frac{1}{N_{\rm stat}}
\sum_{k=1}^{N_{\rm stat}} f^{(k)}.
\label{eq:JK_mean_f}
\end{equation}
The jackknife estimate of the statistical error is then
\begin{equation}
\delta f_{\rm JK}
=
\sqrt{
\frac{N_{\rm stat}-1}{N_{\rm stat}}
\sum_{k=1}^{N_{\rm stat}}
\left(
f^{(k)}-\overline{f}_{\rm JK}
\right)^2
}.
\label{eq:JK_error_f}
\end{equation}
If desired, the leading finite-sample bias associated with the nonlinear transformation $f(Z)\propto \log Z$ can be reduced using the bias-corrected jackknife estimator,
but here we do not implement it.

In the present analysis, Eq.~\eqref{eq:JK_mean_f} is used for the
central value, while Eq.~\eqref{eq:JK_error_f} is used to estimate its
statistical uncertainty.
The above procedure is well defined only when
\be{
%\overline{Z}>0
%\qquad\text{and}\qquad
Z^{(k)}>0
\quad
\text{for all }k.
\label{eq:JK_positivity}
}
%for sufficient statistics.
If one or more leave-one-out estimates are nonpositive, the
corresponding logarithms are not defined. Discarding such jackknife
replicates would change the resampling distribution and could bias the
result, in particular leading to an underestimate of the statistical
uncertainty. The occurrence of nonpositive $Z^{(k)}$ should therefore
be regarded as a sign that the fluctuations of the partition-function
estimator are too large, or that $N_{\rm stat}$ is insufficient for a
reliable determination of the free energy density.

\bibliography{bibliography}

\end{document}